\documentclass{article}

\usepackage{arxiv}

\usepackage[utf8]{inputenc} 
\usepackage[T1]{fontenc}    
\usepackage{hyperref}       
\usepackage{url}            
\usepackage{booktabs}       
\usepackage{amsfonts}       
\usepackage{nicefrac}       
\usepackage{microtype}      
\usepackage{lipsum}
\usepackage[font=small,labelfont=bf]{caption} 

\usepackage{graphicx,calc}
\usepackage{amsmath}
\usepackage[english]{babel}
\usepackage{eurosym}
\usepackage{stmaryrd}

\graphicspath{{IMGs//}}

\title{Disarrangements and instabilities in augmented 1D hyperelasticity}

\author{
  S. Palumbo \\
  Department of Civil, Environmental and Mechanical Engineering, University of Trento, Italy\\
  \texttt{stefania.palumbo@unitn.it} \\
  \And
  L. Deseri \\
  Department of Civil, Environmental and Mechanical Engineering, University of Trento, Italy\\
  Department of Mechanical Engineering \& Department of Civil and Environmental Engineering,\\ Carnegie Mellon University, Pittsburgh,    PA, USA\\
  \texttt{luca.deseri@unitn.it}\\
  \And
   D. R. Owen\\
   Department of Mathematical Sciences and Center for Nonlinear Analysis, Carnegie Mellon University,\\ Pittsburgh, PA, USA\\
   \texttt{do04@andrew.cmu.edu}\\
   \And
  M. Fraldi\\
  Department of Structures for Engineering and Architecture, University of Napoli Federico II, Italy\\
  Institute for Applied Sciences and Intelligent Systems, National Research Council of Italy\\
  \texttt{fraldi@unina.it} 
}

\begin{document}
\maketitle

\begin{abstract}
In the present work, the overall nonlinear elastic behavior of a 1D multi-modular structure incorporating possible imperfections at the discrete (micro-scale) level, is derived with respect to both tensile and compressive applied loads. The model is built up through the repetition of $n$ units, each one comprising two rigid rods having equal lengths, linked by means of pointwise constraints capable to elastically limit motions in terms of relative translations (sliders) and rotations (hinges). 
The mechanical response of the structure is analyzed by varying the number $n$ of the elemental moduli, as well as in the limit case of infinite number of infinitesimal constituents, in light of the theory of (first order) Structured Deformations (SDs), that interprets the deformation of any continuum body as the projection, at the macroscopic scale, of geometrical changes occurring at the level of its sub-macroscopic elements. In this way, a wide family of nonlinear elastic behaviors is generated by tuning internal microstructural parameters, the tensile buckling and the classical Euler's Elastica under compressive loads resulting as special cases in the so-called \textit{continuum limit} --say when $n\rightarrow\infty$. Finally, by plotting the results in terms of first Piola-Kirchhoff stress versus macroscopic stretch, it is for the first time demonstrated that such SDs-based 1D models can be helpfully used to generalize some standard hyperelastic behaviors by additionally taking into account instability phenomena and concealed defects.
\end{abstract}


\section{Introduction}
Classic Continuum Mechanics represents a standard, consolidated instrument to describe the macroscopic response of a wide class of materials under several loading conditions \cite{Cowin_2007,Fraldi_2013}. Its basic assumption of uniform material distribution, stress and strain fields within an infinitesimal neighborhood of each material point, by definition, makes however it inadequate for special cases in which material heterogeneities at the microscale, defects and/or effects of underlying microstuctures influence the overall response of the media and hence necessitate an adequate expression at the macroscopic level.\\
By following analytical approaches or heuristic methods, homogenization theory \cite{nematnasser,FRALDI_2004} is generally invoked to obtain functional relationships among overall continuum fields of interest (i.e. macro-stress, macro-strain and elastic moduli) and microstructural parameters and properties of micro-constituents of the so-called Representative Volume Elements (RVE). Nevertheless, standard homogenization theories do not allow to take into account any possible kinematics that the body undergoes at different scales, that are instead exhibited in some man-made materials and in biological matter characterized by hierarchical architectures. Selected examples of continua which would require more detailed description of their underlying microstructure are auxetic foams \cite{lakes,evans}, soft composites whose mechanical response depends on local buckling phenomena \cite{bertoldi2010,Ruocco_2012} as well as nested tensegrity systems met in a vast number of biomaterials, from the cytoskeleton of human cells up to tissues \cite{Ingber2014,Ingber_2003,Stamenovic,PALUMBO2018}.
A possible way to describe the macroscopic results of microstructural kinematics in complex continua is to adopt the theory of (first order) Structured Deformations (SDs) \cite{Del_Piero_1993,Owen_2004,Deseri2002,stableeq,Deseri2014,DeseriOwen2015}. This allows the incorporation of the effects of sub-macroscopic \textit{piecewise-smooth} geometrical changes that appear \textit{smooth} at the macroscopic level, by \textit{de facto} formulating an improved theory of elasticity with \textit{space-like disarrangements}, such as slips or formation of voids, in which what happens at the body's macroscopic scale is the result of deformations occurring at a lower level.\\
Within this framework, the present work provides a first paradigm of SDs-based 1D mechanical model, incorporating disarrangements, defects, compressive and tensile buckling mechanisms \cite{Zaccaria_2011} at the local level\footnote{Note that a recent paper by Caddemi et al. \cite{Caddemi2015} analyzed compressive and tensile buckling effects in the Euler-Bernoulli column characterized by the presence of sole internal elastic sliders to model shear deformation singularities, by exploiting distribution theory and involving elastic deformation of the bars.}, that finally result in a complex hyperelastic (reversible) behavior which could be properly modulated by prescribed microstructural parameters. It is demonstrated that, with the sole weapon of few elemental degrees of freedom, instability phenomena and several nonlinear elastic behaviors, commonly observed at the macroscopic level, can be all obtained --in analytical way-- as a result of ruling mechanisms concealed at lower scales.\\
On the basis of the theoretical outcomes, it is also expected that such SDs-based models could be usefully exploited for conceiving new non-standard materials and for properly coupling elasticity, growth and remodeling in biological structures \cite{Lubarda_2002,Nappi_2015,FRALDI2018}, including some experimentally recognized \textit{on/off} and threshold-induced switching mechanisms exhibited by biomaterials in response to mechanical stimuli.
\section{Fundamentals of Structured Deformations theory}
\label{SDth}
A (first order) Structured Deformation (SD) \cite{Del_Piero_1993,Ravello} of a body occupying a region $\mathcal{A}$ of an Euclidean space $\mathcal{E}$ with translation space $\mathcal{V}$ is defined by the pair $\left(\textbf{g},\textbf{G}\right) $ of (sufficiently) smooth fields $\textbf{g}:\mathcal{A}\rightarrow\mathcal{E}$, which represents the \textit{macroscopic deformation}, and $\textbf{G}:\mathcal{A}\rightarrow Lin\mathcal{V}$, named \textit{deformation without disarrangements}, as it would coincide with the gradient of $\textbf{g}$ if no disarrangements occurred. In the hypothesis that the vector mapping $\textbf{g}$ is injective and that it and the tensor field $\textbf{G}$ satisfy the condition --known as accomodation inequality-- that there exists a positive number $m$ such that
\begin{equation}
\label{accineq}
m<\det\textbf{G}\left( \textbf{x}_0\right) \leq \det\nabla\textbf{g}\left( \textbf{x}_0\right) \qquad  \forall \,\, \textbf{x}_0\in\mathcal{A}\,,
\end{equation}
the approximation theorem \cite{Del_Piero_1993} assures that it is possible to find (at least) a sequence $\left\lbrace \textbf{g}_n\right\rbrace $ of piecewise-smooth and injective functions --called approximating (or determining) sequences-- defined on $\mathcal{A}$ such that     
\begin{equation}
\label{appth}
\textbf{g}:=\lim_{n \to \infty} \textbf{g}_n \,, \quad \textbf{G}:=\lim_{n \to \infty} \nabla\textbf{g}_n \,,
\end{equation}
in the sense of uniform convergence, $L^{\infty}$, where $\nabla(\bullet)=\partial(\bullet)/\partial\textbf{x}_0=(\bullet)\otimes\nabla$ represents the gradient of a vector function, being $\nabla$ the nabla operator and $\otimes$ the dyadic product. A relaxed, $L^{1}$, convergence has also been shown to hold, thereby proving that $\textbf{g}$ and $\textbf{G}$ are obtainable as volume averages of $\textbf{g}_n$ and $\nabla\textbf{g}_n$, respectively. Thus, according to the basic principle of the SDs theory, the smooth deformations detectable at the macroscopic level can be interpreted as the result of a limit operation from the submacroscopic scale. There, on the contrary, discontinuities such as slips and separations --referred to as \textit{disarrangements}-- are permitted for the determining sequences. Because the limit $\textbf{G}$ in \eqref{appth}$_{2}$ does not need to represent the gradient of any deformation and, hence, it differs from the classic gradient of the macroscopic deformation $\textbf{g}$ in \eqref{appth}$_{1}$, the tensor 
\begin{equation}
\label{distens}
\textbf{M}:=\nabla\textbf{g}-\textbf{G}
\end{equation} 
is introduced to account for the deformation amount relative to disarrangements, accordingly named \textit{deformation due to disarrangements} or disarrangements tensor. It is worth noting that a very revealing identification relation is available for such tensor in terms of the discontinuities $\llbracket \textbf{g}_n \rrbracket$ of the determining sequences introduced above, namely
\begin{equation}
\textbf{M}=\lim_{\delta \to 0}\lim_{n \to \infty} \dfrac{1}{\emph{vol}\mathcal{B}\left( \textbf{x}_0;\delta\right)} \int_{\Gamma\left( \textbf{g}_n\right) \cap \mathcal{B}\left( \textbf{x}_0;\delta\right)} \llbracket \textbf{g}_n \rrbracket\left( \textbf{y}_0 \right) \otimes \boldsymbol{\nu} \left( \textbf{y}_0 \right) \mathrm{d}A_{\textbf{y}_0}\,,
\label{M}
\end{equation}
where $\mathcal{B}\left( \textbf{x}_0;\delta\right)$ is a fixed ball of radius $\delta$ centered at $\textbf{x}_0\in\mathcal{A}$, $\boldsymbol{\nu} \left( \textbf{y}_0 \right)$ is the unit normal to the jump set $\Gamma\left( \textbf{g}_n\right)$ at a point $\textbf{y}_0\in\Gamma\left( \textbf{g}_n\right)$, while $\emph{vol}\mathcal{B}\left( \textbf{x}_0;\delta\right)$ is the volume of the ball previously introduced. The fact that here \textit{first order} disarrangements as averages of jumps on the approximating functions $\textbf{g}_n$ are considered, while possible measures of the jumps on the gradients of such functions are not introduced in the analysis of the geometry of the continuum, allows to identify the pair $(\textbf{g},\textbf{G})$ as \textit{first order} SD.

\subsection{Factorization of a SD}
\label{apxfact}
\begin{figure} [h!]
\centering
\includegraphics[width=.8 \textwidth]{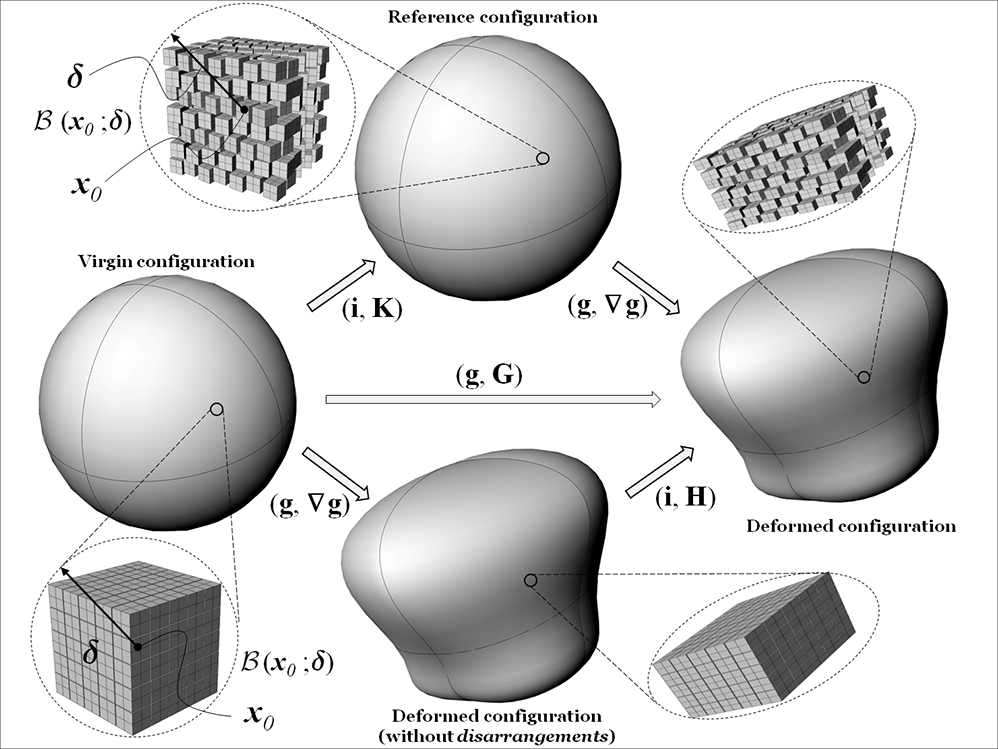} 
\caption{Sketch of the kinematics of a SD decomposed according its two possible factorizations \eqref{factorization1} and \eqref{factorization2}.}
\label{fig.configurations}
\end{figure}

Two (first order) SDs can be composed to give a third SD of the same kind according to the following rule \cite{Del_Piero_1993}:
\begin{equation}
\label{SDcomp}
(\tilde{\textbf{g}},\tilde{\textbf{G}})\circ(\textbf{g},\textbf{G}):=(\tilde{\textbf{g}}\circ\textbf{g},(\tilde{\textbf{G}}\circ\textbf{g})\textbf{G}),
\end{equation}
where $\circ$ denotes the operation of composition. From this definition, it follows that any SD $(\textbf{g},\textbf{G})$ can be factorized as
\begin{equation}
\label{factorization1}
(\textbf{g},\textbf{G})=(\textbf{g},\nabla\textbf{g})\circ(\textbf{i},\textbf{K})\,,
\end{equation}
with $\textbf{i}(\textbf{x}_0):=\textbf{x}_0$ representing the identity mapping and $\textbf{K}:=(\nabla\textbf{g})^{-1}\textbf{G}$, for all $\textbf{x}_0\in\mathcal{A}$. This means that any given (first order) SD can be obtained as the succession of a purely sub-macroscopic SD, namely $(\textbf{i},\textbf{K})$, that maps the body from the \textit{virgin} to the \textit{reference configuration}, and a classic (purely macroscopic) deformation, namely $(\textbf{g},\nabla\textbf{g})$, that maps the body from the \textit{reference} to the \textit{deformed configuration} (see Figure \ref{fig.configurations}). Therefore, a further --virgin-- configuration is needed to be added to the well known reference and deformed ones of the classic Continuum Mechanics, in order to take into account the distinction between the body before and after a deformation at the sub-macroscopic scale, not detectable macroscopically. An equivalent factorization reading as
\begin{equation}
\label{factorization2}
(\textbf{g},\textbf{G})=(\textbf{i},\textbf{H})\circ(\textbf{g},\nabla\textbf{g})\,,
\end{equation}   
being $\textbf{H}:=(\textbf{G}(\nabla\textbf{g})^{-1})\circ\textbf{g}^{-1}$, can be further performed, which allows to interpret the general (first order) SD $(\textbf{g},\textbf{G})$ as the composition of a classic deformation, again $(\textbf{g},\nabla\textbf{g})$, that maps the body from the virgin configuration to the \textit{deformed configuration without disarrangements}, and a purely sub-macroscopic deformation, namely $(\textbf{i},\textbf{H})$, that maps the body from the deformed configuration without disarrangements to the deformed configuration (see Figure \ref{fig.configurations}). In observance of the condition \eqref{accineq}, it follows that $0<\det\textbf{K}=\det\textbf{H}\leq1$.
\subsection{Decomposition of stresses and constitutive assumptions}
\label{apxdec}
Balance laws (of forces and momenta) for a body undergoing a (first order) SD can be written in terms of a proper stress measure. The factorization \eqref{factorization1} and the addition of the virgin configuration require a refinement of such laws in order to take into account the presence of non-classical deformations. Specifically, the refined balance of forces expressed with reference to the virgin configuration reads as follows:
\begin{equation}
\nabla\cdot\left(\textbf{S} \textbf{K}^*\right)-\textbf{S} \nabla\cdot\textbf{K}^*+\nabla\textbf{S}\left[ \left( \det\textbf{K}\right) \textbf{I}-\textbf{K}^*\right]+\textbf{b}_v=\textbf{0}\,,
\end{equation}
where $\textbf{S}$ is the traditional first Piola-Kirchhoff stress tensor, $\textbf{K}^*:=\left(\det\textbf{K}\right)\textbf{K}^{-T}$ is the adjugate of $\textbf{K}$ and $\textbf{b}_v:=\left( \det\textbf{K}\right)\textbf{b}_r$ is the body force per unit volume in the virgin configuration, being $\textbf{b}_r$ the one in the reference configuration.
By virtue of the identification \cite{Ravello,Owen_2004,Owen_1998} of the term $\nabla\cdot\left(\textbf{S}\textbf{K}^*\right)$ with the volume density of total contact forces without disarrangements and of the term $-\textbf{S} \nabla\cdot\textbf{K}^*+\nabla\textbf{S}\left[ \left( \det\textbf{K}\right) \textbf{I}-\textbf{K}^*\right]$ with the volume density of total contact forces due to disarrangements, the quantity 
\begin{equation}
\textbf{S}_{\backslash}:=\textbf{S}\textbf{K}^*,
\label{S-no-disarrangements}
\end{equation}
that can be also interpreted as the stress relative to the virgin configuration, is defined as \textit{stress without disarrangements}, while 
\begin{equation}
\textbf{S}_d:=\left( \det\textbf{K}\right) \textbf{S}-\textbf{S}_{\backslash}
\label{S-due-to-disarrangements}
\end{equation}
is named \textit{stress due to disarrangements}. The resulting additive decomposition of the stress:
\begin{equation}
\label{stressdec}
\left( \det\textbf{K}\right) \textbf{S}=\textbf{S}_{\backslash}+\textbf{S}_d
\end{equation}
has been proved to be unique and universal \cite{Deseri_2015} and provides that the two amounts of stress --$\textbf{S}_{\backslash}$ and $\textbf{S}_d$-- are related through the \textit{consistency relation}
\begin{equation}
\textbf{S}_{\backslash}\left( \textbf{K}^T-\textbf{I}\right) =\textbf{S}_d\,,
\label{consistency}
\end{equation}
where $\textbf{I}$ represents the second-order identity tensor.\\
By taking into account that the SDs theory permits energy to be stored by means of both smooth and non-smooth sub-macroscopic geometrical changes, a free energy function can be assigned as dependent on a pair formed by any combination of kinematic tensors chosen among $\textbf{G}$, $\textbf{M}$ and $\nabla\textbf{g}$ \cite{Deseri_2003}. From an operational point of view, as pointed out in detail in \cite{Deseri_2003} and \cite{Owen-2017}, there are essentially two possible ways to analyze undergoing SDs within a body. A first one is to select a constitutive class, by prescribing constitutive equations for the stresses $\textbf{S}_{\backslash}$ and $\textbf{S}_d$ based on a chosen free energy, then using the relation \eqref{consistency} to restrict the class of admissible processes for the system. A second way is instead to select a constitutive class by directly introducing a stress-strain law involving the total (Piola-Kirchhoff) stress $\textbf{S}$, again based on a chosen free energy, and calculate $\textbf{S}_{\backslash}$ and $\textbf{S}_d$ through the definitions \eqref{S-no-disarrangements} and \eqref{S-due-to-disarrangements}, respectively. This way to proceed does not put any restriction on the kinematical processes and identically verifies the consistency relation \eqref{consistency}.\\
As an illustration of the first procedure, by writing the free energy in the form $\Psi(\textbf{G},\textbf{M})$, the following constitutive assumptions can be made:
\begin{gather}
\label{constclass1}
\textbf{S}_{\backslash}=\left( \det\textbf{K}\right) D_{\textbf{G}}\Psi(\textbf{G},\textbf{M})\,, \\
\label{constclass2}
\textbf{S}_d=\left( \det\textbf{K}\right)D_{\textbf{M}}\Psi(\textbf{G},\textbf{M})\,,
\end{gather}
where $D_{\textbf{G}}$ and $D_{\textbf{M}}$ indicate the partial derivatives with respect to $\textbf{G}$ and $\textbf{M}$, respectively. The choice of the constitutive relations \eqref{constclass1} and \eqref{constclass2} reflects the identification of the stresses $\textbf{S}_{\backslash}$ and $\textbf{S}_d$ as "driving tractions" associated to the tensors of the deformation without disarrangements and of the deformation due to disarrangements, respectively. By referring to the literature for an extensive discussion (\cite{DeseriOwen2015}, \cite{Deseri_2003}), it is here important to recall that he constitutive relations \eqref{constclass1} and \eqref{constclass2} provide, by substitution into \eqref{stressdec}, the total stress $\textbf{S}$ in the following form:
\begin{equation}
\textbf{S}=D_{\textbf{G}}\Psi(\textbf{G},\textbf{M})+D_{\textbf{M}}\Psi(\textbf{G},\textbf{M}).
\label{Stress-constitutive}
\end{equation}
This form clearly highlights that the free energy is \textit{de facto} a generalized potential of the stress, as one would expect for a generalized hyperelastic material; in absence of disarrangements --that is when $\textbf{M}=\textbf{0}$, $\textbf{K}=\textbf{I}$ and $\textbf{G}\equiv\nabla\textbf{g}$-- it is immediate to see that \eqref{Stress-constitutive} returns the classical equation for standard hyperelastic continua, provided that $D_{\textbf{M}}\Psi(\textbf{G},\textbf{0})\equiv\textbf{0}$.\\
As an illustration of the the second procedure, starting from the same free energy $\Psi(\textbf{G},\textbf{M})$ as above, one may assume directly the constitutive relation \eqref{Stress-constitutive} for $\textbf{S}$ and calculate from it, through eqs \eqref{S-no-disarrangements} and \eqref{S-due-to-disarrangements}, the stresses $\textbf{S}_{\backslash}$ and $\textbf{S}_d$ that, in general, will differ from \eqref{constclass1} and \eqref{constclass2}. In this second procedure, the consistency relation \eqref{consistency} is identically satisfied in all motions of the body and, hence, does not provide a tensorial equation restricting the pair $(\textbf{G},\textbf{M})$ or, equivalently, the pair $(\textbf{G},\nabla\textbf{g})$. In this case, alternative restrictions can be provided through specific choices of determining sequences (e.g. see \eqref{sequence-tensile} in the examples below) and through a requirement of equilibrium at sub-macroscopic levels (e.g. see eq. \eqref{force}).
\section{Analysis of the 1D multi-modular structure}
In what follows, the nonlinear elastic response of a multi-modular structure under tensile and compressive loads is analyzed, incorporating tensile and compressive buckling and possible imperfections at the discrete (micro-scale) level.
\begin{figure} [!h]
\centering
\includegraphics[width=.6 \textwidth]{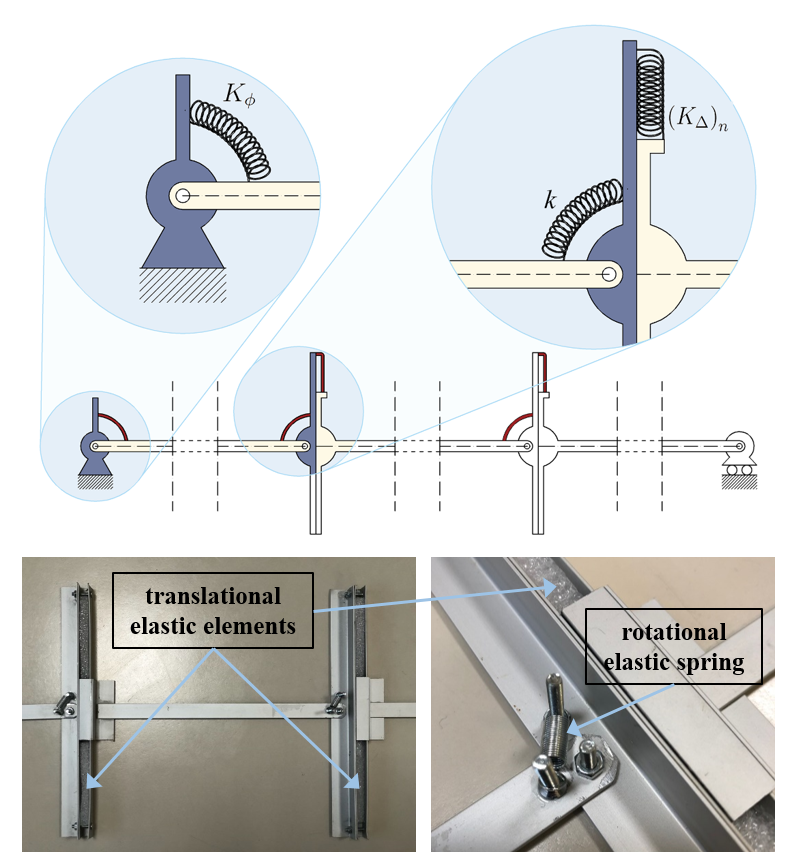} 
\caption{Sketch of the 1D periodic structure at the sub-macroscopic scale, with possible details of lateral hinge (at the left end of the system) equipped with a rotational elastic spring having stiffness $K_{\phi}$ and of the internal constraint able to respond both as a slider, enriched with a spring of stiffness $(K_{\Delta})_n$, and as a hinge, equipped with a rotational spring of stiffness $k$. The pictures at the bottom show details of the actually realized multi-modular system: bars and sliders are made by standard aluminium profiles available in commerce; the hinges are endowed with elastic springs while the elastic elements of the sliders are obtained by embedding in them expanded poly-ethylene (EPE) rods.}
\label{fig.constraint_det}
\end{figure}
The entire 1D structure results from the repetition of $n$ units each one comprising two rigid rods having equal lengths, linked by means of pointwise elastic constraints (see Figure \ref{fig.constraint_det}). At the extremities, the whole system is anchored to a hinge equipped with a rotational spring, to the left end, and by a roller on the right one, where either tensile and compressive external dead loads can be applied. More in detail, under proper constitutive assumptions, it is found that --in case of tension (Figure \ref{fig.1D_structure}\hyperref[fig.1D_structure]{A})-- the elastic hinges do not activate relative rotations and the intermediate constraints react only as sliders, by so replicating --as the number $n$ of units tends to infinity and under prescribed constitutive assumptions for the springs-- the behavior of the elementary single-degree-of-freedom system presented in \cite{Zaccaria_2011}, the latter constituting the first example of structure undergoing buckling under tensile dead load. In the complementary situation, i.e. when a compressive load is applied (Figure \ref{fig.1D_structure}\hyperref[fig.1D_structure]{B}), only the elastic hinges interconnecting the two parts of each unit are enabled to respond, the sliders remaining dormant and, as $n\rightarrow\infty$, the model giving back the classical buckled \textit{Elastica}. In both the cases of tension and compression --reproduced by the authors through a toy system in Figure \ref{fig.experiment}-- the deformation of the multi-modular structure has been described by means of SDs, outlining a first 1D paradigm for this theory.
\begin{figure} [!h]
\centering
\includegraphics[width=1 \textwidth]{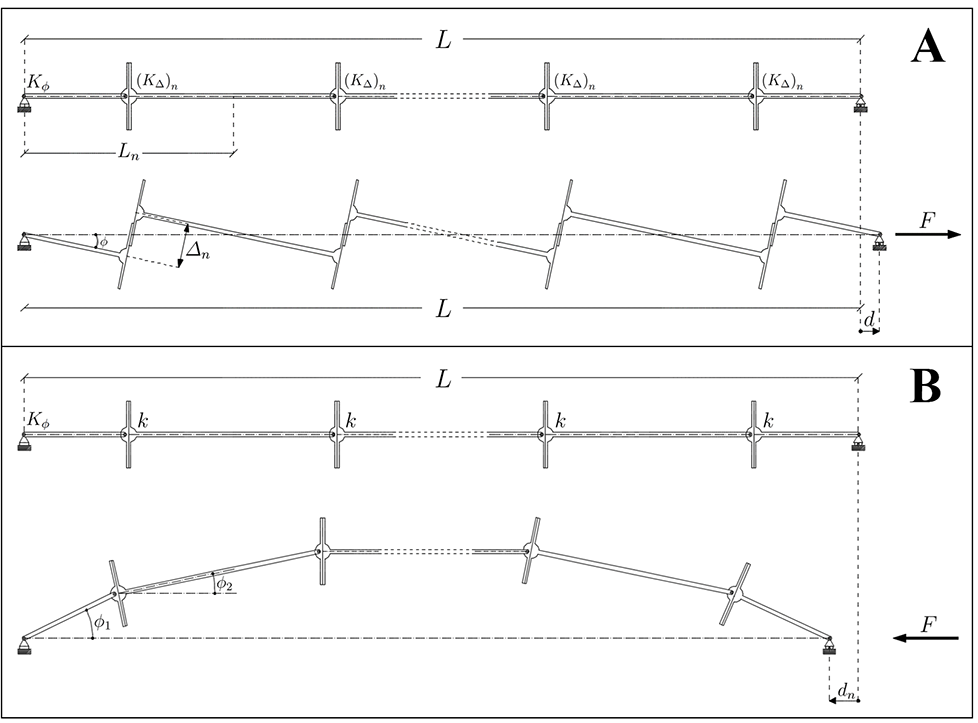} 
\caption{Sketch of the 1D underformed periodic structure and of its deformed configuration, observed at a sub-macroscopic scale, both under \textbf{A)} tensile and \textbf{B)} compressive dead loads. The structure, having whole length $L$, is made up of elemental moduli each having length $L_n$ and consisting of two rigid rods interconnected by means of an internal constraint responding as a slider, with a spring of stiffness $(K_{\Delta})_n$, under tension and as a hinge, with a rotational spring of stiffness $k$, under compression. A hinge --with a rotational spring of stiffness $K_{\phi}$-- and a roller bound the structure respectively at its left and right ends, the latter being the point of application of the external load $F$. In the  deformed configuration under tensile load, $\Delta_n$ represents the sliding between the endpoints of each slider, $\phi$ the rotation angle and $d$ the displacement of the left end. In the deformed configuration under compressive load, $\phi_i$ represents the $i$-th rotation angle and $d_n$ the displacement of the left end.}
\label{fig.1D_structure}
\end{figure}

\begin{figure} [!h]
\centering
\includegraphics[width=.7 \textwidth]{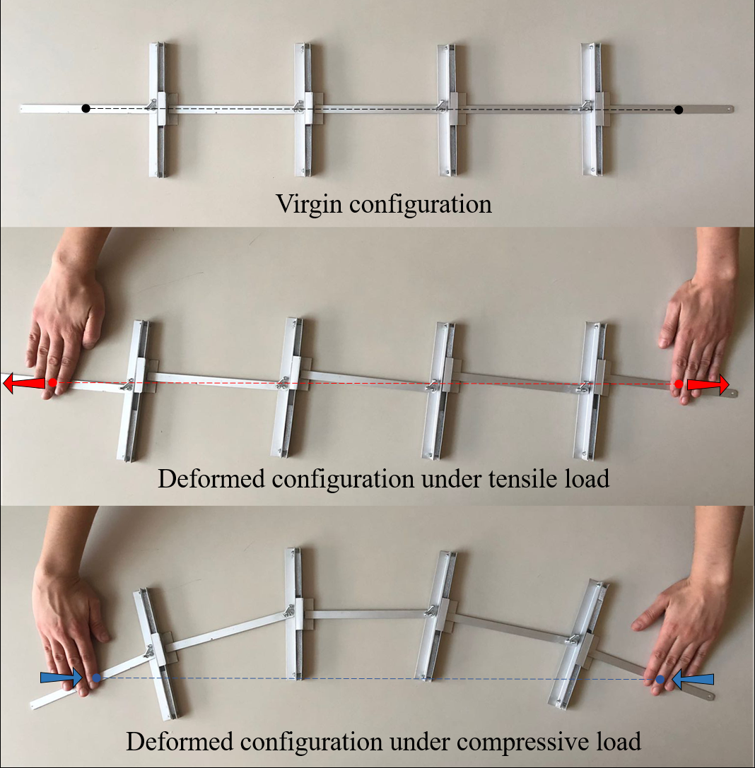} 
\caption{Prototype of the multi-modular structure under exam realized with aluminium bars, roughly loaded by tensile and compressive forces applied at the system ends. Note that, in both the cases, the sliders' orientations and the bars' slopes describe deformed configurations very close to those predicted by the proposed theoretical model.}
\label{fig.experiment}
\end{figure}

\subsection{Bifurcation modes for the elementary system}
\label{secn1}
\begin{figure} [!h]
\centering
\includegraphics[width=.95 \textwidth]{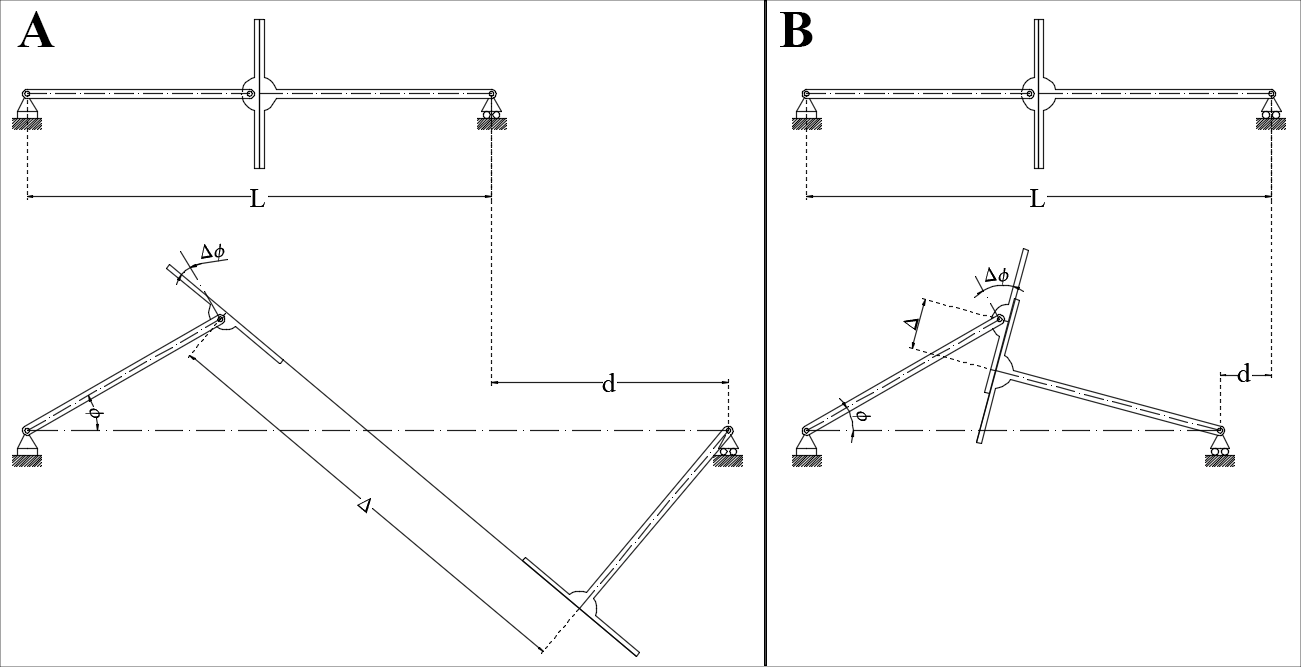} 
\caption{Sketch of the 1D underformed elementary (single-unit) structure and of its possible buckled configuration both under \textbf{A)} tensile and \textbf{B)} compressive dead loads.}
\label{fig.module}
\end{figure}
In the present section, the response of the structure's elemental unit (i.e. for $n=1$), under both tensile and compressive dead loads, is briefly examined. In principle, by assuming finite value of stiffness for all the elastic springs comprising the elementary system, a kinematics involving the contemporary opening of the slider and rotation of the central hinge would be admissible, both under traction and compression, as exemplified in Figure \ref{fig.module}. Therefore, by firstly considering the case of traction (see Figure \ref{fig.module}\hyperref[fig.module]{A}), the most general expression of the internal energy $U$ --that combines the elastic contributions of the transverse spring $K_{\Delta}$ stretching for a tract $\Delta$, of the internal rotational one $k$ which rotates of an angle $\Delta\phi$ and, finally, of the lateral spring $K_{\phi}$ undergoing a rotation $\phi$ with respect to the undeformed condition-- can be given as      
\begin{equation}
\label{intenU}
U=\dfrac{1}{2}\left[K_{\phi}\phi^2+k\Delta\phi^2+K_{\Delta}\Delta^2\right],\quad \Delta=\dfrac{L}{2}\left[\tan\left(\phi+\Delta\phi\right)+\sin\phi\sec\left(\phi+\Delta\phi\right)\right],
\end{equation}
$L$ being the length of the single-unit system. Then, the total potential energy reads
\begin{equation}
\label{potenW}
W=U-Fd, \quad d=\dfrac{L}{2}\left[\left(1+\cos\Delta\phi\right)\sec\left(\phi+\Delta\phi\right)-2\right],  
\end{equation}
herein $d$ being the displacement of the system's left end. It can be verified that the solution of the linearized equilibrium problem --standardly formulated by adopting the stationary total potential energy principle with reference to the kinematics variables $\phi$ and $\Delta\phi$-- provides the following linear relation between the two rotation angles in the first tract of the non-trivial post-buckling path (i.e. for small rotations $\phi$ and $\Delta\phi$):
\begin{equation}
\label{dphitens}
\Delta\phi=\dfrac{-4k+2K_{\phi}+K_{\Delta}L^2 +\sqrt{4\left(4k^2+12k K_{\phi}+K_{\phi}^2\right)+4K_{\Delta}\left(6k+K_{\phi}\right)L^2+K_{\Delta}^2L^4}}{8k}\phi
\end{equation}
and a critical tensile load reading as
\begin{equation}
\label{fcrt}
F_{cr}=\dfrac{-4k-2K_{\phi}+K_{\Delta}L^2 +\sqrt{4\left(4k^2+12k K_{\phi}+K_{\phi}^2\right)+4K_{\Delta}\left(6k+K_{\phi}\right)L^2+K_{\Delta}^2L^4}}{4L}.
\end{equation}
An analogous procedure can be followed to study the buckling of the elementary structure under compression. In this case, the general expressions \eqref{intenU}$_1$ and \eqref{potenW}$_1$ for the internal and total potential energy, respectively, can be adopted with reference to the possible kinematics depicted in Figure \ref{fig.module}\hyperref[fig.module]{B}, by writing
\begin{equation}
\label{deltac}
\Delta=\dfrac{L}{2}\sec\left(\Delta\phi-\phi\right)\left[\sin\left(\Delta\phi-\phi\right)-\sin\phi\right]
\end{equation}
and
\begin{equation}
\label{dc}
d=L\left[1-\cos\left(\dfrac{\Delta\phi}{2}\right)\cos\left(\dfrac{3}{2}\Delta\phi-2\phi\right)\sec\left(\Delta\phi-\phi\right)\right].
\end{equation}
By solving the linearized equilibrium problem, one again finds a linear law relating the two rotations in the first bifurcation mode, that is
\begin{equation}
\label{dphicomp}
\Delta\phi=\dfrac{3\left(4k-2K_{\phi}-K_{\Delta}L^2\right)-\sqrt{4\left(36k^2+28kK_{\phi}+9K_{\phi}^2\right)+4K_{\Delta}\left(K_{\phi}-2k\right)L^2+K_{\Delta}^2L^4}}{2\left(8k-K_{\Delta} L^2\right)}\phi,
\end{equation}  
associated to the lowest compressive critical force given by
\begin{equation}
\label{fcrcomp}
F_{cr}=\dfrac{12k+6K_{\phi}+K_{\Delta}L^2-\sqrt{4\left(36k^2+28kK_{\phi}+9K_{\phi}^2\right)+4K_{\Delta}\left(K_{\phi}-2k\right)L^2+K_{\Delta}^2L^4}}{4L}.
\end{equation} 
The above results show how the critical loads and the specific bifurcation mode depend on the concurrent participation of all the spring-components according to their elastic constants, with both $\Delta$ and $\Delta\phi$ potentially contemporary non-vanishing. However, in such general conditions, it is not possible to find closed form solutions of the not-linearized equilibrium problems --either in the simplest case of the elementary (single-unit) structure-- and, moreover, the number of independent kinematics variable would grow as $2n$ while increasing the number of constituents moduli, thus resulting not manageable without the aid of numerical approaches already for not too large values of $n$. On these bases, to the aim of studying analytically the mechanical response of the proposed 1D multi-modular architecture and to then provide a description in light of the SDs theory, we assume, in the present work, specific relationships among the magnitudes of the springs' elastic constants, that guarantee the onset of bifurcation modes exhibiting expedient symmetries. In particular, under the specific assumption that the internal rotational springs behave as infinitely rigid, say $k\rightarrow \infty$ with respect to the elastic constants of the other springs, the tensile loading condition yields the post-critical deformation dynamics illustrated in Figure \ref{fig.1D_structure}\hyperref[fig.1D_structure]{A}, with the sole opening sliders and frozen hinges, thus periodically reproducing the deformation response of the elementary paradigm presented in \cite{Zaccaria_2011} for tensile buckling. Indeed, it is possible to not hardly verify that the central hinge rotation $\Delta\phi$ in \eqref{dphitens} --found for the unitary system-- results vanishing for any value of the angle $\phi$ when making $k$ divergent, while the displacement $d$ in \eqref{potenW}$_2$, the sliders' relative sliding $\Delta$ in \eqref{intenU}$_2$ and the tensile critical force $F_{cr}$ in \eqref{fcrt} assuming the same expressions that can be found in the next section \ref{sectraction} --for $n=1$-- by postulating \textit{ab initio} the non-participation of the internal hinges to the structure's elastic response under tension. On the other hand, the hypothesis that the sliders exhibit infinite stiffness $K_{\Delta}$ if compared to the one of the rotational springs, provides, under compression, the bifurcation mode shown in Figure \ref{fig.1D_structure}\hyperref[fig.1D_structure]{B} for an exemplifying multi-modular structure, in which the sole internal relative rotations take place, without any sliding. For the simplest system examined in this section, one can find that the sliders' spring extension $\Delta$ in \eqref{deltac} actually vanishes when $K_{\Delta}\rightarrow \infty$, while the left end displacement in \eqref{dc}, the rods relative rotation $\Delta\phi$ in \eqref{dphicomp} and, finally, the critical compressive load in \eqref{fcrcomp} coincide with the ones provided in section \ref{sec:comp} --where the sliders are assumed to be rigid-- when considering $n=1$.   

\subsection{1D multi-modular structure under tensile load}
\label{sectraction}
\begin{figure}[h]
\centering
\includegraphics[width=1 \textwidth]{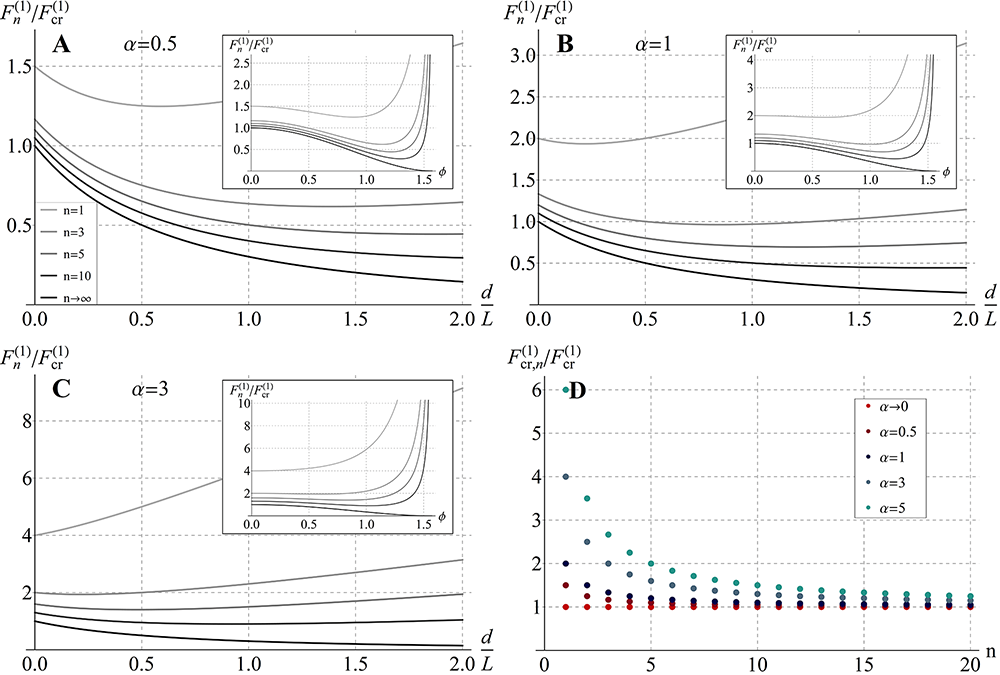} 
\caption{Tensile force $F_n^{(1)}$ --normalized with respect to the (limit) critical load $F_{cr}^{(1)}$-- as function of the boundary displacement normalized to the length of the whole structure $d/L$ and of the rotation $\phi$ (in the insets) for increasing number of elemental moduli ($n=1,3,5,10$) up to the continuum limit $n\rightarrow\infty$, for different values of the ratio $\alpha$ between the translational and rotational springs stiffnesses: \textbf{A)} $\alpha=0.5$, \textbf{B)} $\alpha=1$ and \textbf{C)} $\alpha=3$. \textbf{D)} Normalized tensile critical load $F_{cr,n}^{(1)}/F_{cr}^{(1)}$ as function of $n$, plotted for different values of $\alpha$ ($\alpha\rightarrow0,\alpha=0.5,1,3,5$).}
\label{fig.FphiFcr1}
\end{figure}

\begin{figure} [h]
\centering
\includegraphics[width=1 \textwidth]{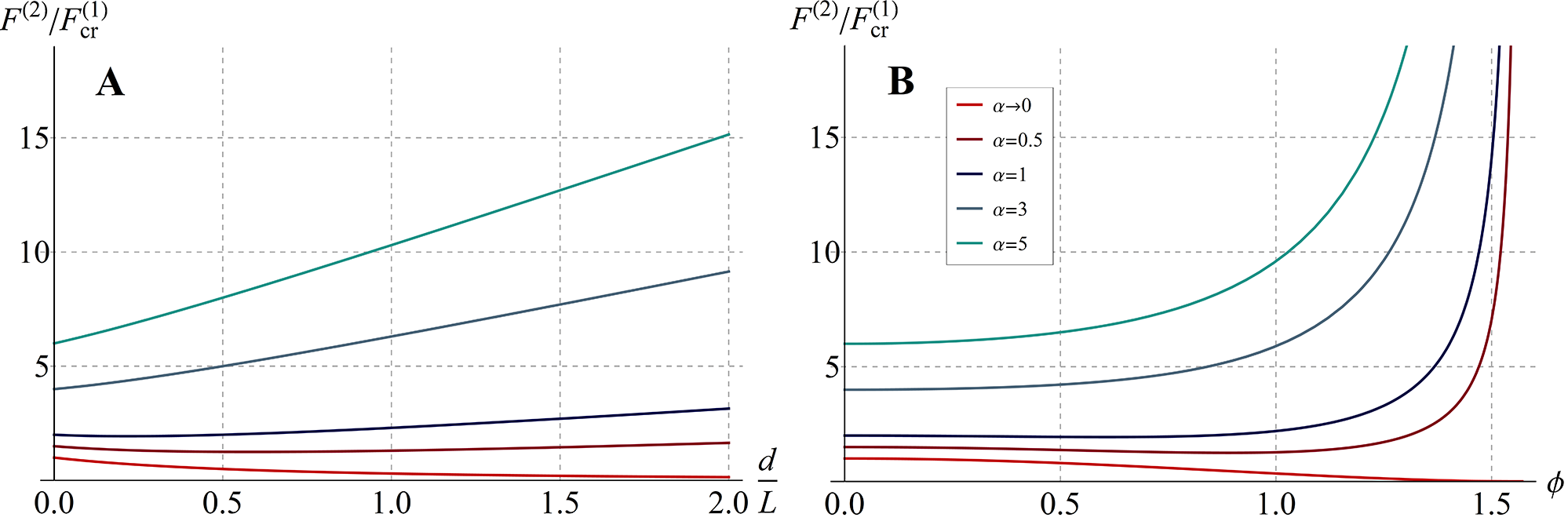} 
\caption{Tensile force $F^{(2)}$ --normalized with respect to the limit critical load $F_{cr}^{(1)}$-- as function \textbf{A)} of the normalized displacement $d/L$ and \textbf{B)} of the rotation $\phi$, for different values of the ratio $\alpha$ between the translational and rotational springs stiffnesses ($\alpha\rightarrow0,\alpha=0.5,1,3,5$).}
\label{fig.Fphi2}
\end{figure}

\begin{figure} [h!]
\centering
\includegraphics[width=1 \textwidth]{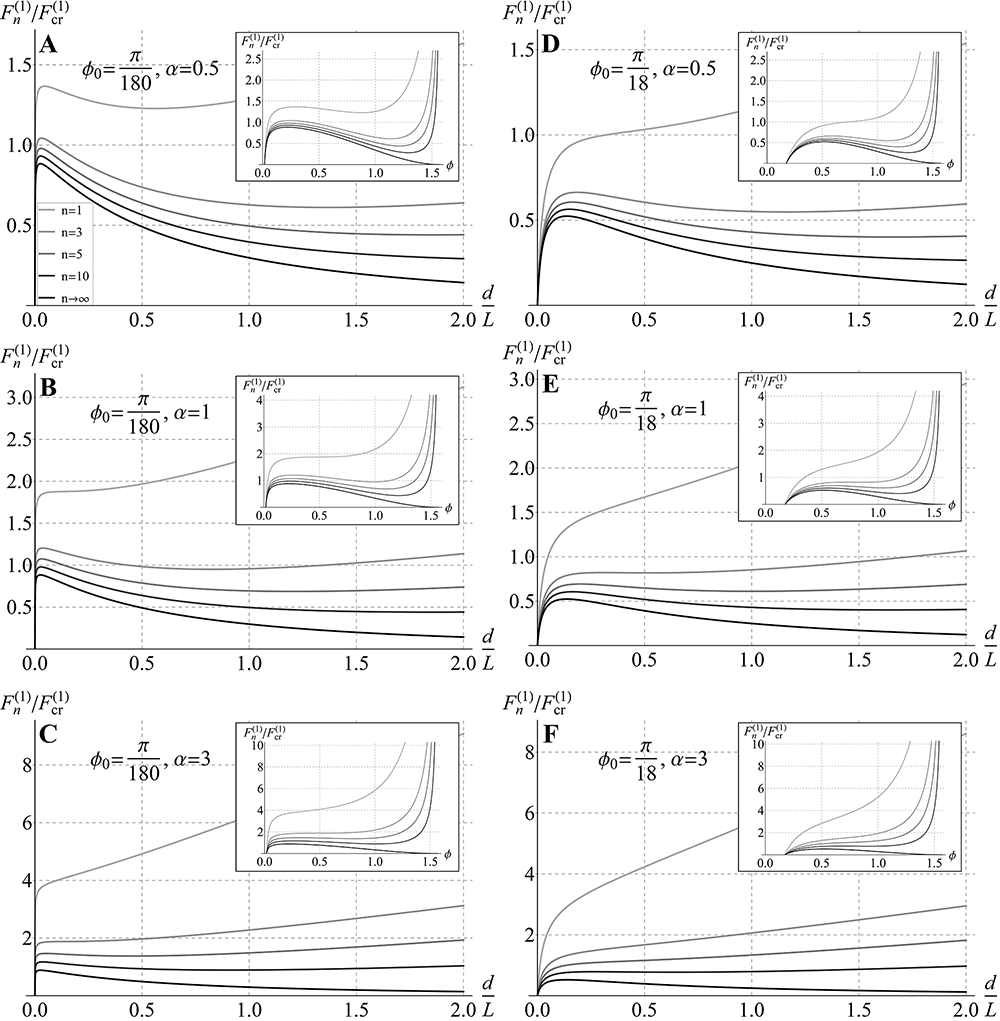} 
\caption{Tensile force $F_n^{(1)}$ --normalized with respect to the limit critical load $F_{cr}^{(1)}$-- against normalized displacement $d/L$ and rotation $\phi$ (in the insets) for increasing number of elemental moduli composing the structure ($n=1,3,5,10$) up to the continuum limit $n\rightarrow\infty$, for different values of the imperfection $\phi_0$ and of the ratio $\alpha$ between translational and rotational springs stiffnesses: \textbf{A)} $\phi_0=\pi/180$, $\alpha=0.5$, \textbf{B)} $\phi_0=\pi/180$, $\alpha=1$, \textbf{C)} $\phi_0=\pi/180$, $\alpha=3$, \textbf{D)} $\phi_0=\pi/18$, $\alpha=0.5$, \textbf{E)} $\phi_0=\pi/18$, $\alpha=1$, \textbf{F)} $\phi_0=\pi/18$, $\alpha=3$.}
\label{fig.F1tensimperf}
\end{figure}

\begin{figure} [h]
\centering
\includegraphics[width=1 \textwidth]{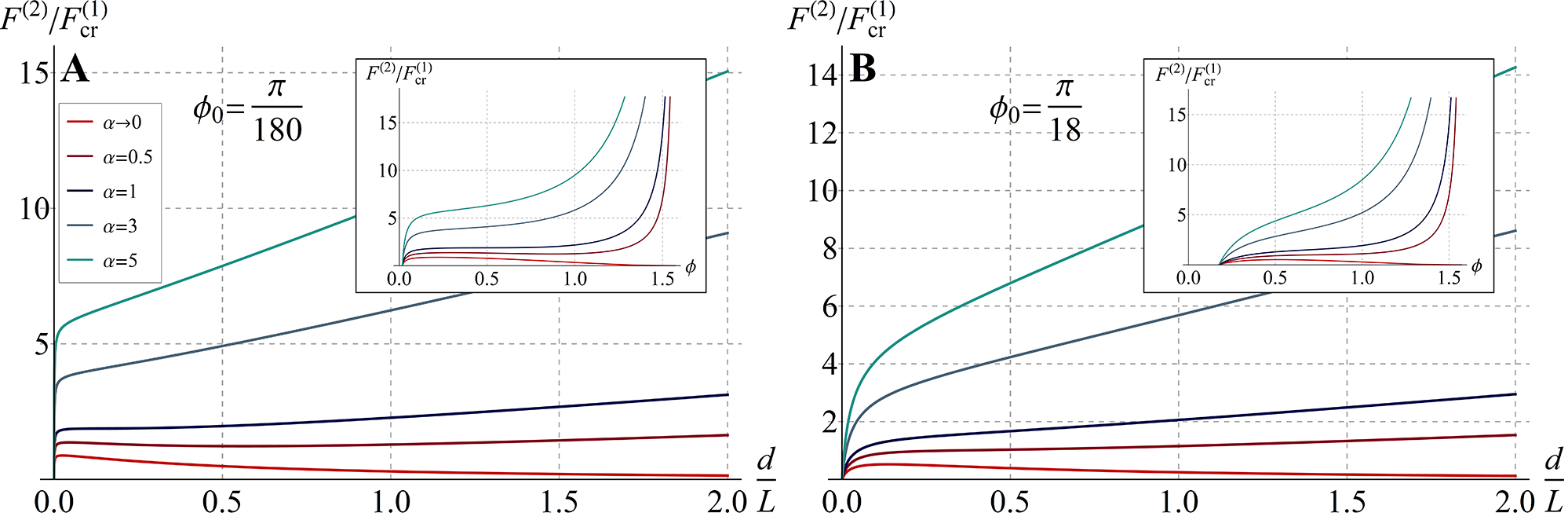}
\caption{Tensile force $F^{(2)}$ --normalized with respect to the limit critical load $F_{cr}^{(1)}$-- versus normalized displacement $d/L$ and rotation $\phi$ (in the insets) for different values of the ratio $\alpha$ between translational and rotational springs stiffnesses ($\alpha\rightarrow0,\alpha=0.5,1,3,5$) and values of the imperfection $\phi_0$: \textbf{A)} $\phi_0=\pi/180$ and \textbf{B)} $\phi_0=\pi/18$.}
\label{fig.F2tensimperf}
\end{figure}

Let us consider a structure made of any number $n$ of modular elements undergoing tensile dead load. With reference to the bifurcation mode illustrated in Figure \ref{fig.1D_structure}\hyperref[fig.1D_structure]{A}, provided by infinitely stiff internal rotational springs, geometrical arguments lead to the conclusion that the whole compatible kinematics is ruled by the sole rotational degree of freedom represented by the rotation $\phi$. As a consequence, the internal elastic energy --that the deforming structure stores through the sliders translational springs and the lateral rotational one-- is given by
\begin{equation}
\label{intenertens}
U_n(\phi)=\dfrac{1}{2}\left[ K_{\phi} \phi^2+n\left( K_{\Delta}\right)_n \Delta_n^2\right],
\end{equation}
where $\Delta_n=L_n\tan\phi$ is the relative sliding between the two adjacent endpoints of each slider, $L_n=L/n$ being the length of each modular element and $L$ the length of the whole structure, $K_{\phi}$ is the stiffness of the rotational spring associated to the hinge on the left end of the structure, while $(K_{\Delta})_n$ is the stiffness associated to each one of the $n$ internal transverse springs. Then, the total potential energy, formed by the internal energy of the whole system minus the work done by the external load, can be written as follows
\begin{equation}
W_n(\phi)=U_n(\phi)-F d(\phi)\,,
\end{equation}
where $F$ is the applied tensile load and $d(\phi)=L\left( \sec\phi-1\right)$ is the corresponding displacement at the right end, independent of the number $n$ of constituent moduli. Consequently, by making the total potential energy stationary, the solutions of the equilibrium problem are two, say the trivial solution $\phi=0, \forall F$, and a non-trivial one, characterized by the following expression of the force, $F_n$, in principle depending on the number of units $n$:
\begin{equation}
\label{force}
\dfrac{\partial W_n(\phi)}{\partial\phi}=0 \Rightarrow F_n= \dfrac{K_{\phi}\phi\cos\phi}{L\tan\phi}+\dfrac{\left( K_{\Delta}\right)_n L}{n\cos\phi}\,,
\end{equation}
where the first term coincides with the solution obtained for the structure in \cite{Zaccaria_2011}, while the second one is due to the presence of the intermediate transverse springs. It is worth noting that $\phi=0$ is a bifurcation path for equilibria. Indeed, upon evaluating \eqref{force} as $\phi \rightarrow 0$, one immediately finds the critical value $F_{cr,n}$ of the force as
\begin{equation}
\label{fcrtens}
F_{cr,n}= \dfrac{K_{\phi}}{L}+\dfrac{\left( K_{\Delta}\right)_n L}{n}\,.
\end{equation}
As a matter of fact, the behavior of the system strictly depends on how the stiffness $\left( K_{\Delta}\right)_n$ scales with the number of units $n$. However, it is easy to verify that --by assuming that the sliders' stiffness scales by a power law, i.e. $\propto n^p, p\in \mathbb{R}^+$-- internal energy and forces at the equilibrium assume finite values for $0\leq p \leq 1$, giving divergent results as $p>1$. Therefore, without loss of generality, we analyze the two cases in which $p=0$ and $p=1$. More specifically, if one assumes that the sliders' stiffness is constant with $n$, that is $\left( K_{\Delta}\right)_n=K_{\Delta}$, the quantities in eqs \eqref{intenertens}, \eqref{force} and \eqref{fcrtens} depend on $n$ and converge respectively to the energy, the force and the critical load of the structure in \cite{Zaccaria_2011} as $n\rightarrow\infty$, since the contribution of the internal springs converges to zero, i.e.:
\begin{equation}
\label{case1}
\begin{split}
U^{(1)}&:=\lim_{n\rightarrow\infty}U_n^{(1)}=\lim_{n\rightarrow\infty}U_n\left(\left( K_{\Delta}\right)_n\rightarrow K_{\Delta}\right)=\lim_{n\rightarrow\infty}\dfrac{1}{2}\left( K_{\phi} \phi^2+ \dfrac{K_{\Delta} L^2 \tan^2\phi}{n} \right)=\dfrac{1}{2} K_{\phi} \phi^2,\\
F^{(1)}&:=\lim_{n\rightarrow\infty}F_n^{(1)}=\lim_{n\rightarrow\infty}F_n\left(\left( K_{\Delta}\right)_n\rightarrow K_{\Delta}\right)=\lim_{n\rightarrow\infty}\dfrac{K_{\phi}\phi\cos\phi}{L\tan\phi}+\dfrac{K_{\Delta} L}{n\cos\phi}=\dfrac{K_{\phi}\phi\cos\phi}{L\tan\phi},\\
F_{cr}^{(1)}&:=\lim_{n\rightarrow\infty}F_{cr,n}^{(1)}=\lim_{n\rightarrow\infty}F_{cr,n}\left(\left( K_{\Delta}\right)_n\rightarrow K_{\Delta}\right)=\lim_{n\rightarrow\infty}\dfrac{K_{\phi}}{L}+\dfrac{ K_{\Delta} L}{n}=\dfrac{K_{\phi}}{L}.
\end{split}
\end{equation}
Upon assuming a different scaling between $\left( K_{\Delta}\right)_n$ and $n$, say $\left( K_{\Delta}\right)_n=n K_{\Delta}$, i.e. a stiffness proportional to the number of modular units composing the structure, the quantities in eqs \eqref{intenertens}, \eqref{force} and \eqref{fcrtens} turn out to be independent of $n$:
\begin{equation}
\label{case2}
\begin{split}
U^{(2)} &:=\lim_{n\rightarrow\infty} U_n^{(2)}=\lim_{n\rightarrow\infty}U_n\left(\left( K_{\Delta}\right)_n\rightarrow nK_{\Delta}\right)=\dfrac{1}{2}\left( K_{\phi} \phi^2+ K_{\Delta} L^2 \tan^2\phi \right) ,\\
F^{(2)} &:= \lim_{n\rightarrow\infty} F_n^{(2)}=\lim_{n\rightarrow\infty} F_n\left(\left( K_{\Delta}\right)_n\rightarrow nK_{\Delta}\right)=\dfrac{K_{\phi}\phi\cos\phi}{L\tan\phi}+\dfrac{K_{\Delta} L}{\cos\phi},\\
F_{cr}^{(2)} &:=\lim_{n\rightarrow\infty} F_{cr,n}^{(2)}=\lim_{n\rightarrow\infty}F_{cr,n}\left(\left( K_{\Delta}\right)_n\rightarrow nK_{\Delta}\right)=\dfrac{K_{\phi}}{L}+ K_{\Delta} L.
\end{split}
\end{equation}
In both the cases, the response of the system depends on the relationship between the values of stiffnesses $K_{\phi}$ and $K_{\Delta}$, therefore a coefficient $\alpha$ is suitably introduced so that $K_{\Delta}=\alpha K_{\phi}/L^2$. It is found that, while the trivial solution is always stable up to the critical load and unstable after that value, the non-trivial post-critical behavior depends on $\alpha$. Figure \ref{fig.FphiFcr1} shows the force $F_n^{(1)}$ --normalized with respect to the (limit) critical load $F_{cr}^{(1)}$-- both as function of the normalized displacement $d/L$ (or overall engineering strain) and of the rotation angle $\phi$, for different values of $\alpha$ and $n$. In particular, for the addressed case of $\left( K_{\Delta}\right)_n=K_{\Delta}$, analysis of the second derivative of the strain energy provides that the non-trivial post-bifurcation path is stable under the condition $\alpha>(5/3)\,n$ for any rotation angle $\phi$ (by way of example, see the curve obtained for $\alpha=3$ and $n=1$ in Figure \ref{fig.FphiFcr1}\hyperref[fig.FphiFcr1]{C}). Otherwise, as also detectable from Figures \ref{fig.FphiFcr1}\hyperref[fig.FphiFcr1]{A,B,C}, the system undertakes an unstable non-trivial post-critical behavior, that can reach stability at some finite deformation depending on the number of moduli $n$ (see the rising tracts after initial softening in the insets of Figure \ref{fig.FphiFcr1}), with the exception of the limit case $n\rightarrow\infty$, for which the structure remains unstable over the whole deformation range. An analogous behavior is found when $\left( K_{\Delta}\right)_n=nK_{\Delta}$, in this case the discriminating condition being given by $\alpha>5/3$, as observable in Figure \ref{fig.Fphi2}, that shows the trend of the force $F^{(2)}\equiv F_n^{(2)}$ --again normalized with respect to the (limit) critical load $F_{cr}^{(1)}$-- as function of both the above defined normalized displacement $d/L$ and the rotation angle $\phi$, for different values of $\alpha$.  In Figure \ref{fig.FphiFcr1}\hyperref[fig.FphiFcr1]{D}, the trend of the normalized critical load $F_{cr,n}^{(1)}/F_{cr}^{(1)}=(n+\alpha)/n$ as function of $n$, for different values of the ratio $\alpha$, is also displayed, while the critical load $F_{cr}^{(2)}$ is simply given by $F_{cr}^{(2)}=(1+\alpha)F_{cr}^{(1)}$, its expression thus coinciding with the previous case when $\alpha\rightarrow0$.
\\
In analogy to what has been done in \cite{Zaccaria_2011}, it is of interest to evaluate the response of the system in the presence of an \textit{imperfection}, i.e. a defect of the structure at the microscopic level, for instance assumed to be an initial inclination $\phi_0$ of the rods. In this case, the total potential energy can be written as follows:
\begin{equation}
W(\phi;\phi_0)=\dfrac{1}{2} \left[ K_{\phi} (\phi-\phi_0)^2+\dfrac{\left( K_{\Delta}\right)_n L^2}{n} (\tan\phi-\tan\phi_0)^2\right]-FL\left( \sec\phi-\sec\phi_0\right),
\end{equation}
consequently giving the force-rotation equilibrium relationship as
\begin{equation}
\label{fnimperf}
F_n= \dfrac{K_{\phi}(\phi-\phi_0)\cos\phi}{L\tan\phi}+\dfrac{\left( K_{\Delta}\right)_n L}{n} \left( \dfrac{1}{\cos\phi}-\dfrac{\tan\phi_0}{\sin\phi}\right), 
\end{equation}
so that eqs \eqref{case1}$_2$ 
and \eqref{case2}$_2$ can be now respectively replaced with: 
\begin{gather}
F_n^{(1)}= \dfrac{K_{\phi}(\phi-\phi_0)\cos\phi}{L\tan\phi}+\dfrac{K_{\Delta} L}{n} \left( \dfrac{1}{\cos\phi}-\dfrac{\tan\phi_0}{\sin\phi}\right) \quad \xrightarrow[n \rightarrow \infty]{} \quad F^{(1)}=\dfrac{K_{\phi}(\phi-\phi_0)\cos\phi}{L\tan\phi} \\
\text{and}\quad
F^{(2)} \equiv F_n^{(2)}=\dfrac{K_{\phi}(\phi-\phi_0)\cos\phi}{L\tan\phi}+K_{\Delta} L \left( \dfrac{1}{\cos\phi}-\dfrac{\tan\phi_0}{\sin\phi}\right).
\end{gather}
Figure \ref{fig.F1tensimperf} shows the normalized force $F_n^{(1)}/F_{cr}^{(1)}$ as function of the normalized displacement $d/L$ --for the present case being $d=L(\sec \phi - \sec \phi_0)$-- and of the rotation $\phi$, for different values of $n$ and $\alpha$, as well as for two initial values of the imperfection $\phi_0$. Similarly, Figure \ref{fig.F2tensimperf} shows the trend of $F^{(2)}/F_{cr}^{(1)}$ as function of $d/L$ and $\phi$, for several $\alpha$ and for the same values of the initial imperfections.

\subsubsection{SDs-based formulation for tensile loads}
The kinematics that the system experiences when it is subjected to tensile load, specifically in the considered non-trivial post-buckling phase (Figure \ref{fig.1D_structure}\hyperref[fig.1D_structure]{A}), can be suitably described by means of the (first order) SDs theory. In particular, the generic material point of the studied body in the virgin configuration (in which, by definition, it is neither macroscopically nor sub-macroscopically deformed) can be identified by a position vector $\textbf{x}_0=x_0\hat{\textbf{e}}_1$, with $x_0 \in \left[ 0,L\right] $ and $\hat{\textbf{e}}_i$, $i\in\left\lbrace 1,2\right\rbrace$, indicating the unit vectors of two-dimensional Cartesian coordinate system.  The approximating functions, that map each point $\textbf{x}_0$ to the deformed configuration (in which the body is both macroscopically and sub-macroscopically deformed) at the sub-macroscopic scale, can be expressed as  
\begin{equation}
\textbf{g}_n^h \left( \textbf{x}_0 \right)=\textbf{x}_0+\textbf{u}_n^h \left( \textbf{x}_0 \right),
\end{equation}
where, by virtue of geometrical arguments, the displacement $\textbf{u}_n^h$ reads
\begin{multline}
\textbf{u}_n^h \left( \textbf{x}_0 \right)=\left[ x_0 \left( \cos\phi-1\right)+hL_n\left( \sec\phi-\cos\phi\right) \right] \hat{\textbf{e}}_1 - \left[ \left( x_0-hL_n\right) \sin\phi\right] \hat{\textbf{e}}_2  \\
\forall\,\,x_0 \, \in \, \Biggl] \frac{L_n}{2}\left( 2h-1 +\delta\left( h\right) \right) , \frac{L_n}{2}\left( 2h+1-\delta\left( h-n \right) \right) \Biggr[\,\,,
\label{sequence-tensile}
\end{multline}
where $h\in\left\lbrace 0,1,..,n\right\rbrace $ is a translation index and $\delta(h-p)$ the Kronecker delta function taking the value $1$ when $h=p$ and $0$ otherwise. Then, the deformation gradient associated to the approximating sequence takes the form
\begin{equation}
\label{Ggradg}
\nabla\textbf{g}_n^h = \textbf{I}+\nabla\textbf{u}_n^h=\cos\phi\,\hat{\textbf{e}}_1\otimes\hat{\textbf{e}}_1-\sin\phi\,\hat{\textbf{e}}_2\otimes\hat{\textbf{e}}_1+\hat{\textbf{e}}_2\otimes\hat{\textbf{e}}_2 \equiv \textbf{G}\,,
\end{equation}
which, since not depending on $n$, naturally provides also the deformation without disarrangements $\textbf{G}$ (see eq.\eqref{appth}$_{2}$). Herein, $\textbf{I}=\hat{\textbf{e}}_1\otimes\hat{\textbf{e}}_1+\hat{\textbf{e}}_2\otimes\hat{\textbf{e}}_2$ is the identity tensor. On the other hand, $\textbf{g}_n^h(\textbf{x}_0)$ uniformly converges to the macroscopic deformation $\textbf{g}(\textbf{x}_0)=\textbf{x}_0+\textbf{u}(\textbf{x}_0)$, with
\begin{equation}
\label{macrodef}
\textbf{u}(\textbf{x}_0)=\lim_{n \to \infty} \textbf{u}_n^h(\textbf{x}_0)=\left( \sec\phi-1\right) \textbf{x}_0\,,
\end{equation}
and, consequently, the related classical deformation gradient is 
\begin{equation}
\label{gradgtrac}
\nabla\textbf{g}=\textbf{I}+\nabla\textbf{u}=\sec\phi\,\hat{\textbf{e}}_1\otimes\hat{\textbf{e}}_1+\hat{\textbf{e}}_2\otimes\hat{\textbf{e}}_2\,.
\end{equation}
It is worth highlighting that the tensor $\textbf{G}$ obtained in eq.\eqref{Ggradg} satisfies the requirement of inextensibility of the single constituent rods --a direct consequence of the assumption of rigidity-- as $\vert\textbf{G}\hat{\textbf{e}}_1\vert=1$. Additionally, the result of the geometry above entails $\textbf{G}\neq \nabla\textbf{g}$, the studied 1D model thus revealing the presence of sub-macroscopic \textit{disarrangements}. As a matter of fact, at the macro-scale, the non-classical nature of the deformation considered for the system under study turns out to produce a threshold-activated pure axial elongation whose magnitude is ruled by the disarrangement degree at the sub-macroscopic level (see eq. \eqref{macrodef}), for all the values of the angle $\phi$ in the range of interest $\left[ 0,\pi/2\right[$. The corresponding overall constitutive response can be deduced by plotting force \textit{versus} displacement as shown in Figure \ref{fig.FphiFcr1} for the limiting case $n\rightarrow \infty$ and in Figure \ref{fig.Fphi2} as a function of the choice made about the stiffness of the internal sliders springs. Also, it has to be highlighted that --within the whole range-- the fields $\textbf{g}$ and $\textbf{G}$ satisfy the so-called accomodation inequality \eqref{accineq}.\\
The presence of disarrangements is taken into account by the disarrangements tensor, that in this case is given by
\begin{equation}
\label{Mtrac}
\textbf{M}=\sin\phi\left( \tan\phi \,\hat{\textbf{e}}_1\otimes\hat{\textbf{e}}_1+\hat{\textbf{e}}_2\otimes\hat{\textbf{e}}_1\right),
\end{equation}
while the tensor $\textbf{K}$, related to the purely sub-macroscopic deformation, takes the form
\begin{equation}
\label{Ktrac}
\textbf{K}=\cos^2\phi \,\hat{\textbf{e}}_1\otimes\hat{\textbf{e}}_1-\sin\phi\,\hat{\textbf{e}}_2\otimes\hat{\textbf{e}}_1\,+\hat{\textbf{e}}_2\otimes\hat{\textbf{e}}_2\,.
\end{equation}
Also, by exploiting the one-dimensional nature of the stress regime related to the uni-axial applied external load, the equilibrium allows one to write the first Piola-Kirchhoff stress tensor in the form
\begin{equation}
\label{Sint}
\textbf{S}=\frac{F}{A} \, \hat{\textbf{e}}_1\otimes\hat{\textbf{e}}_1\,,
\end{equation}
where $A$ is the transverse area of the structure in the reference configuration and $F$ is given by $F^{(1)}$ or $F^{(2)}$, depending upon the choice made on $\left( K_{\Delta}\right)_n$. Thus, according to their definitions, the stress without disarrangements and the stress due to disarrangements turn out to be respectively 
\begin{gather}
\label{Snodis}
\textbf{S}_{\backslash}=\frac{F}{A} \left( \hat{\textbf{e}}_1\otimes\hat{\textbf{e}}_1 + \sin\phi \,\hat{\textbf{e}}_1\otimes\hat{\textbf{e}}_2 \right)\\
\label{Ssidis}
\text{and}\quad \textbf{S}_d=-\frac{F}{A} \sin\phi \left( \sin\phi\,\hat{\textbf{e}}_1\otimes\hat{\textbf{e}}_1 +  \hat{\textbf{e}}_1\otimes\hat{\textbf{e}}_2 \right).
\end{gather}
Finally, from the kinematical point of view, the SD describing the deformation of the system when characterized by an imperfection $\phi_0$, namely $(\textbf{z},\textbf{Z})$, can be seen as the composition of two SDs according to the definition \eqref{SDcomp}, that is:
\begin{equation}
(\textbf{z},\textbf{Z}) = (\textbf{g},\textbf{G})\circ (\textbf{g}_0,\textbf{G}_0)^{-1} = \left(\textbf{g}\circ\textbf{g}_0^{-1},\textbf{G} \textbf{G}_0^{-1}\right) ,
\end{equation}
being, by virtue of the invertibility of the tensor $\textbf{G}$ in the range of interest for $\phi$, $(\textbf{g}_0,\textbf{G}_0)=(\textbf{g},\textbf{G})\vert_{\phi=\phi_0}$ and $(\textbf{g}_0,\textbf{G}_0)^{-1}=(\textbf{g}_0^{-1},\textbf{G}_0^{-1})$  its inverse.

\subsubsection{Consistency of discrete and SDs-based approaches: an argument in support of augmented hyperelasticity}
By regarding the multi-modular structure under study as a 1D continuum and by looking at the limiting quantities obtained through both the relations \eqref{case1} and \eqref{case2} for constant and proportional to $n$ sliders' stiffness, respectively, it appears evident that the limiting energies per unit volume (e.g. divided by the cross sectional area $A$ of the structural arms and the overall length $L$ of the system) are not potentials for the stress, the sole nonzero component $S_{11}=F/A$ being in fact not obtainable as derivative of the energy with respect to any standard strain measure coming from the deformation gradient. This would force one to admit that --at least within the classical framework of continua-- as $n\rightarrow\infty$, the resulting 1D continuum material cannot be thought of as hyperelastic. To be convinced of the need to use SDs, the naturally arising question is then whether or not SDs give a way to find a generalized potential for $\textbf{S}$, for instance in the form \eqref{Stress-constitutive}.\\
With this in mind, we start by remarking that, in the case at hand, the geometry is known, namely both the macroscopic deformation \eqref{macrodef} and the limit of the gradients of its approximating sequence \eqref{Ggradg} have been found, and, furthermore, the stress is statically determined. However, these two pieces of information do not suffice to ensure that the resulting continuum behaves as an augmented (generalized) hyperelastic material, that is the stress obeys the \eqref{Stress-constitutive}. To demonstrate that it is the case, we have in fact to seek a \textit{free energy} $\psi$ such that \textit{i}) it is a generalized potential for the stress $\textbf{S}$ and \textit{ii}) no dissipation is found for the resulting limiting  material (i.e. as $n\rightarrow\infty$).
With reference to the request \textit{ii}), it is in fact worth to recall that, in general, the overall presence of disarrangements neither requires nor rules out the possibility of having \textit{dissipation} during loading \cite{stableeq,Deseri2014,DeseriOwen2015}. Nonetheless, because the underlying parent discrete systems, discussed in the previous Sections, do not exhibit dissipation behaving as purely elastic at the local scale, it is expected that the effective continuum and its resulting constitutive properties do not entail dissipation.\\
In order to establish if \textit{i}) can be fulfilled, upon analyzing the geometrical changes described by \eqref{sequence-tensile} for the tensile case, we first can note that the discrete system undergoing an imposed axial displacement achieves its balance thanks to the change in configuration of each arm, captured through $\nabla\textbf{g}_n^h$ in \eqref{Ggradg} acting on $\hat{\textbf{e}}_1$, caused by pulling above the critical load and to the energy stored within the sliders because of the openings $\Delta_n$ arising at each of such sites due to the rotations. Somehow, in the limit $n\rightarrow\infty$, the effective continuum mimics a body whose internal fibers reorient according to what happens only in the direction $\hat{\textbf{e}}_1$. In the limit, the features just highlighted are in fact embodied in the first invariant (i.e. the trace) of the tensors $\textbf{G} \hat{\textbf{e}}_1\otimes \textbf{G}\hat{\textbf{e}}_1$ and $\textbf{M} \hat{\textbf{e}}_1 \otimes \textbf{M}\hat{\textbf{e}}_1$. This suggests to represent the target effective free energy density per unit volume of the continuum limit in the form
\begin{equation}
\psi=\Psi(\textbf{G}, \textbf{M})=\tilde{\Psi}(tr(\textbf{G} \hat{\textbf{e}}_1\otimes \textbf{G}\hat{\textbf{e}}_1) + tr(\textbf{M} \hat{\textbf{e}}_1 \otimes \textbf{M}\hat{\textbf{e}}_1)).
\label{free_energy}
\end{equation}
From this assumption, by considering the forms of the tensors $\textbf{G}$ and $\textbf{M}$ given in eqs. \eqref{Ggradg} and \eqref{Mtrac} for the problem at hand, one can find: 
\begin{equation}
\label{drivingforces}
\begin{split}
D_\textbf{G}\Psi(\textbf{G}, \textbf{M}) &= 2 \tilde{\Psi}' \textbf{G} \hat{\textbf{e}}_1 \otimes \hat{\textbf{e}}_1=2 \tilde{\Psi}' \left(\cos\phi \, \hat{\textbf{e}}_1 \otimes \hat{\textbf{e}}_1-\sin\phi \, \hat{\textbf{e}}_2\otimes  \hat{\textbf{e}}_1 \right) \quad \text{and}\\
D_\textbf{M}\Psi(\textbf{G}, \textbf{M}) &= 2 \tilde{\Psi}'  \textbf{M} \hat{\textbf{e}}_1 \otimes \hat{\textbf{e}}_1 =2 \tilde{\Psi}'\left[\left(\sec\phi-\cos\phi\right) \hat{\textbf{e}}_1 \otimes \hat{\textbf{e}}_1+\sin\phi \, \hat{\textbf{e}}_2\otimes  \hat{\textbf{e}}_1\right],
\end{split}
\end{equation}
that, by substitution into \eqref{Stress-constitutive}, provide the following expression for the first Piola-Kirchhoff tensor:
\begin{equation}
\textbf{S}=2 \tilde{\Psi}' \sec\phi \, \hat{\textbf{e}}_1 \otimes \hat{\textbf{e}}_1,
\label{Constitutive-S}
\end{equation}
where the apex indicates the derivative of the function with respect to its argument, here equal to $tr(\textbf{G} \hat{\textbf{e}}_1\otimes \textbf{G}\hat{\textbf{e}}_1) + tr(\textbf{M} \hat{\textbf{e}}_1 \otimes \textbf{M}\hat{\textbf{e}}_1)=\sec^2\phi$. This Piola-Kirchhoff stress tensor, derived by proper constitutive assumptions and according to the SDs theory, has to coincide with the one obtained by means of equilibrium arguments, given in eq.\eqref{Sint}. The imposition of this condition leads to the following differential equation for the free energy density:    
\begin{equation}
\tilde{\Psi}'= \frac{F(\phi)}{2A} \cos\phi.
\label{Differential}
\end{equation}
The integration of both the members of this equation with respect to the argument of the unknown function $\tilde{\Psi} (\sec^2\phi)$ would return a general expression for $\tilde{\Psi}$.
As an example, to show the form that $\tilde{\Psi}$ could assume at equilibrium for a particular choice of the structural constitutive parameters, the case in which the stiffness of the sliders springs scale proportionally to $n$, i.e. $(K_\Delta)_n=n K_\Delta$, is taken into account, so that the response of the system is described by the quantities in \eqref{case2}. Also, for sake of simplicity, it is assumed that no lateral rotational spring is present, namely $K_\phi=0$. Under these conditions, the force at equilibrium is given by $F=K_\Delta L \sec\phi$, so that $\tilde{\Psi}'$ turns out to be constant with respect to its argument and the integration of \eqref{Differential} can be easily performed:
\begin{equation}
\label{Free-energy-Response-1}
\tilde{\Psi}'(\sec^2\phi)=\dfrac{K_\Delta L}{2A}, \quad \tilde{\Psi}(\sec^2\phi)=\dfrac{K_\Delta L}{2A} \sec^2\phi + C.
\end{equation}
The value of the constant of integration $C$ is then found by imposing vanishing energy as rotation approaches zero, so that:
\begin{equation}
\label{Free-energy-Response-2}
C=-\dfrac{K_\Delta L}{2A} \quad \Rightarrow \quad \tilde{\Psi}=\dfrac{K_\Delta L}{2A} \tan^2\phi. 
\end{equation}
It is possible to observe that the elastic energy that one obtains by multiplying $\tilde{\Psi}$ in \eqref{Free-energy-Response-2}$_2$ by the volume of the continuum body, $A L$, actually matches the expression of $U^{(2)}$ in \eqref{case2}$_1$ written for $K_\phi=0$, thus revealing the full consistency of the SDs modeling strategy with the discrete approach. It can be verified that such a conclusion is not influenced by the specific choice made about the system's parameters and can be therefore generalized to different cases.Furthermore, for seek of completeness, if interested in retrieving the stress also before it reaches its critical value, one should properly introduce a Lagrangean multiplier for taking into account the rigidity constraint due to the axial inextensibility of the structure. In the present cases this can be trivially determined from the boundary conditions. However, for a detailed discussion about the stress decomposition in presence of reactive components, the reader is referred to \cite{Owen-2017}, eqs $38$ and $39$.\\
It is also worth noticing that, differently from classical elasticity theory where continua store energy during isothermal processes involving smooth finite deformations, the constitutive assumptions employed in the SDs framework allow the body both to store and dissipate energy while undergoing geometrical changes across the scales, due to the presence of disarrangements \cite{stableeq,Deseri2014,DeseriOwen2015,Deseri_2003}. In particular, the second law of thermodynamics requires that any SD satisfies the following \textit{dissipation inequality}:
\begin{equation}
\dot{\psi}\leq \textbf{S}\cdot (\nabla \textbf{g})^\textbf{.},
\end{equation}
which classically claims that the rate of change of the density of Helmoltz free energy does not exceed the density of stress-power (the dot over the variable indicating its time derivative and the dot as superscript denoting time derivative of the whole content in the parentheses). By virtue of the additive decomposition provided in eqs. \eqref{distens} and \eqref{Stress-constitutive}, such inequality reads
\begin{equation}
\dot{\psi}\leq D_{\textbf{G}}\Psi \cdot \dot{\textbf{G}} + D_{\textbf{M}}\Psi \cdot \dot{\textbf{M}} +D_{\textbf{G}}\Psi \cdot \dot{\textbf{M}} +D_{\textbf{M}}\Psi \cdot \dot{\textbf{G}}.
\end{equation}  
It can be recognized that the sum of the first two terms on the right side of this equation recovers the rate of change of the free energy density $\dot{\psi}$. This means that the sum of the two last terms, in the SDs theory referred to as \textit{mixed (stress) power}, represents the rate of work done by the stress components due to disarrangements and those not associated to disarrangements against the reciprocal kinematics counterparts, being the difference between the free energy rate and the stress-power. Therefore, the \textit{internal dissipation} is given by
\begin{equation}
\label{dissipation}
\Upsilon:=\textbf{S}\cdot (\nabla \textbf{g})^\textbf{.}-\dot{\psi}=D_{\textbf{G}}\Psi\cdot \dot{\textbf{M}} +D_{\textbf{M}}\Psi \cdot \dot{\textbf{G}} \geq 0.
\end{equation}    
As a consequence, \eqref{dissipation} yields a decomposition of the stress-power into a non-dissipative part and a non-negative dissipative part:
\begin{equation}
\textbf{S}\cdot (\nabla \textbf{g})^\textbf{.}=\dot{\psi}+\Upsilon.
\end{equation}   
Then, with reference to the above mentioned request \textit{ii}), it can be observed that the specific SD characterizing the multi-modular system under tensile load provides a fully reversible deformation process, because the dissipation $\Upsilon$ vanishes. Indeed, by employing the relations \eqref{drivingforces} in \eqref{dissipation} and by explicitly calculating the rate of change of the kinematic tensors $\textbf{G}$ and $\textbf{M}$ as:
\begin{equation}
\begin{split}
\dot{\textbf{G}} &=  -\dot{\phi}  \, \left(\sin\phi \,\hat{\textbf{e}}_1\otimes\hat{\textbf{e}}_1+\cos\phi\,\hat{\textbf{e}}_2\otimes\hat{\textbf{e}}_1\right) \quad \text{and} \\
\dot{\textbf{M}} &=   \dot{\phi}  \, \left[\sin\phi \left(1+\sec^2\phi\right)\, \hat{\textbf{e}}_1\otimes\hat{\textbf{e}}_1+\cos\phi\,\hat{\textbf{e}}_2\otimes\hat{\textbf{e}}_1\right],
\end{split}
\label{process}
\end{equation}
where it is considered that the unit vectors $\hat{\textbf{e}}_1$ and $\hat{\textbf{e}}_2$ are time-invariant, it is possible to verify that the dissipation $\Upsilon$ vanishes for any rotation $\phi$.

\subsection{1D multi-modular structure under compressive load}
\label{sec:comp}
\begin{figure}[!h]
\centering
\includegraphics[width=1 \textwidth]{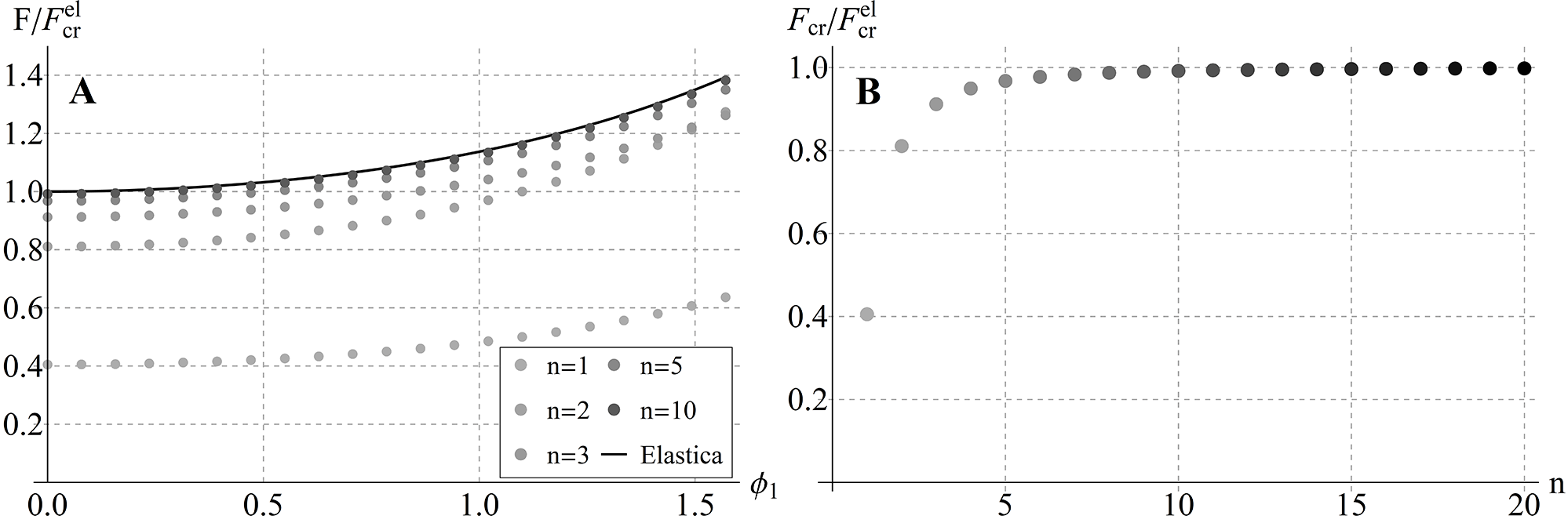}
\caption{\textbf{A)} Compressive load $F$ --normalized with respect to the Elastica critical load $F_{cr}^{el}$ \cite{Bigoni2012}-- as function of the initial rotation $\phi_1=\phi(0)$, for increasing number of elemental moduli composing the structure ($n=1,2,3,5,10$) up to the limit $n\rightarrow\infty$ coinciding with the continuum case of the Elastica. \textbf{B)} Critical compressive load $F_{cr}$ --normalized with respect to the Elastica critical load $F_{cr}^{el}$ \cite{Bigoni2012}-- as function of $n$.}
\label{fig.F_Fcr}
\end{figure}

In this section we analyze the complementary case of a compressive load applied to the examined multi-modular structure, under the hypothesis of infinitely stiff internal transverse springs and, as a consequence, not opening sliders. The activation of the hinges under this type of loading makes possible, in principle, a kinematics characterized by independent rotations for each element, according to the constraints imposed at the endpoints. The effective value of such rotations, i.e. the effective bifurcation mode of the structure, will be then determined by solving the equilibrium problem. Without loss of generality, we will assume that the stiffness of the lateral rotational spring $K_{\phi}$ is much lower than the one of the internal hinges, $k$. In this case, the contribution associated with $K_{\phi}$ to the internal energy stored by the structure while undergoing compression can be neglected. Therefore, by virtue of this consideration and of the assumption of rigidity for the constituent rods, the internal energy can be given as
\begin{equation}
U_n(\phi_1,...,\phi_n)=\frac{1}{2}k \sum_{i=1}^n\left(\phi_i-\phi_{i+1} \right)^2 ,
\end{equation}
where $\phi_i$, $i=\lbrace 1,...,n+1\rbrace$, are the angles that the $n+1$ constituent rods form with respect to the horizontal direction (taken positive when clockwise), with 
\begin{equation}
\sin\phi_{n+1}=-\sin\phi_1-2\sum_{i=2}^n\sin\phi_i
\end{equation}
in order to respect the geometrical constraint on the right endpoint. Then, the total potential energy can be expressed as    
\begin{equation}
W_n(\phi_1,...,\phi_n)=U_n(\phi_1,...,\phi_n)-F d_n(\phi_1,...,\phi_n)\,,
\end{equation}
where
\begin{equation}
d_n(\phi_1,...,\phi_n)=L-L_n\left(\dfrac{\cos\phi_1+\cos\phi_{n+1}}{2} + \sum_{i=2}^n\cos\phi_i\right)
\end{equation}
is the displacement at the boundary and $F$ the compressive external load.\\
To solve the linearized equilibrium problem in case of small rotations, one can first obtain a generalized McLaurin's series expansion of the equilibrium equations $\nabla_{\phi_1,...,\phi_n} W=\textbf{0}$ up to the first order and then solve the system $\left[ \nabla\nabla(W)\vert_{\phi_i=0}\right]\left(\phi_1 ... \phi_n \right)^T =\textbf{0}$, where  $\nabla(\bullet)$ and $\nabla \nabla(\bullet)=\nabla\otimes\nabla(\bullet)$ represent the gradient and the Hessian operators, respectively. One can hence verify that, among all the possible bifurcation modes, the one associated to the lowest critical load is that corresponding to the symmetrical configuration, as exemplified in Figure \ref{fig.1D_structure}\hyperref[fig.1D_structure]{B}. Also, by following physical arguments, the structure that at early stages deforms according to this symmetry is then assumed to preserve it when undergoing large rotations\footnote{Rigorously speaking, this kinematical assumption would neglect the possibility of other  modes (say \textit{zig-zag} as well as localized V-shaped ones) as the deformation grows approaching larger rotations. However, as highlighted in the text, despite these other modes were here not taken into account, numerical calculations confirmed that minimal energy values were only attained by the hypothesized symmetrical kinematics. Nevertheless, the authors do not exclude that, for example in presence of imperfections --such as slight discrepancies among the bar lengths, springs' stiffness etc-- the system could in principle activate not-symmetrical modes more prone to reach lower energy levels.}. Therefore, the total potential energy can be accordingly simplified by considering as independent variables the sole rotations characterizing the left half of the structure, that is   
\begin{equation}
W_n(\phi_1,...,\phi_{a_n})=U_n(\phi_1,...,\phi_{a_n})-F d_n(\phi_1,...,\phi_{a_n}),
\end{equation}
with
\begin{gather}
U_n(\phi_1,...,\phi_{a_n})=k \left[  \sum_{i=1}^{a_n-1}\left(\phi_i-\phi_{i+1} \right)^2 +(2-b_n) \phi_{a_n}^2\right] \\
\text{and} \quad d_n(\phi_1,...,\phi_{a_n})=L-L_n\left( \cos\phi_1+2\sum_{i=2}^{a_n}\cos\phi_i+b_n\right)\,,
\end{gather}
where we set 
\begin{equation}
a_n=\begin{cases}
n/2 \quad &n \, even \\
(n+1)/2 \quad &n \, odd
\end{cases}\quad \text{and} \quad
b_n=\begin{cases}
1 \quad &n \, even \\
0 \quad &n \, odd
\end{cases}\,.
\end{equation}
By explicitly writing the equilibrium equations $\partial_{\phi_i}W=0,\,\forall\,i=\lbrace 1,...,a_n \rbrace$, after some algebraic manipulations, one obtains 
\begin{equation}
\label{discelast}
\begin{cases}
\dfrac{\phi_2-\phi_1}{L_n/2} +\dfrac{F}{k} \sin\phi_1=0 \,\, \qquad &i=1\\
\\
\dfrac{\phi_{i+1}-2\phi_i+\phi_{i-1}}{L_n^2}+\dfrac{F}{kL_n}\sin\phi_i=0 \,\,\qquad &i=2,...,a_n-1\\
\\
\dfrac{\left(b_n-3\right)\phi_{a_n}+\phi_{a_n-1}}{L_n^2}+\dfrac{F}{kL_n} \sin\phi_{a_n}=0 \,\, \qquad &i=a_n
\end{cases}
\end{equation} 
where the discrete version of the problem of the Euler's Elastica under compressive load \cite{Bigoni2012} can be recognized. Specifically, the equations \eqref{discelast}$_{2}$ and \eqref{discelast}$_{3}$ take the place of the well-known second order differential equation
\begin{equation}
\phi^{''}+\frac{F}{B}\sin\phi=0
\end{equation}
and can be exactly identified with it if $n\rightarrow \infty$, by properly setting $k=B/L_n$, where $B$ would represent the bending stiffness of the continuum system. Also, with such a stiffness value, the equation \eqref{discelast}$_{1}$, as $n\rightarrow \infty$, provides the Elastica boundary conditions $\phi^{'}(0)=\phi^{'}(L)=0$.\\
As a matter of fact, one finds that the curves $F$ \textit{versus} $\phi_1$,  obtained by solving the equilibrium problem \eqref{discelast}, quickly converge to the Elastica solution as $n$ increases. Indeed, by substituting $k=B/L_n$, from equations \eqref{discelast}$_{1}$ and \eqref{discelast}$_{2}$, it is possible to derive each rotation as a function of both the angle at the origin $\phi_1=\phi(0)$ and the external force, that is
\begin{equation}
\phi_2=\phi_1-\dfrac{FL_n^2}{2B}\sin\phi_1 \quad \text{and} \quad \phi_{i+1}=2\phi_i-\phi_{i-1}-\dfrac{FL_n^2}{B}\sin\phi_i\,,i=2,...,a_n-1\,.
\end{equation} 
The substitution of these expressions into \eqref{discelast}$_{3}$ provides the equation, here numerically solved, that relates $\phi_1$ and $F$, finally finding the result shown in Figure \ref{fig.F_Fcr}\hyperref[fig.F_Fcr]{A}. Furthermore, by solving the linearized equilibrium problem, one obtains the trend of the bifurcation loads as function of $n$ (see Figure \ref{fig.F_Fcr}\hyperref[fig.F_Fcr]{B}).   

\subsubsection{SDs-based formulation for compressive loads}
\begin{figure}[!h]
\centering
\includegraphics[width=1 \textwidth]{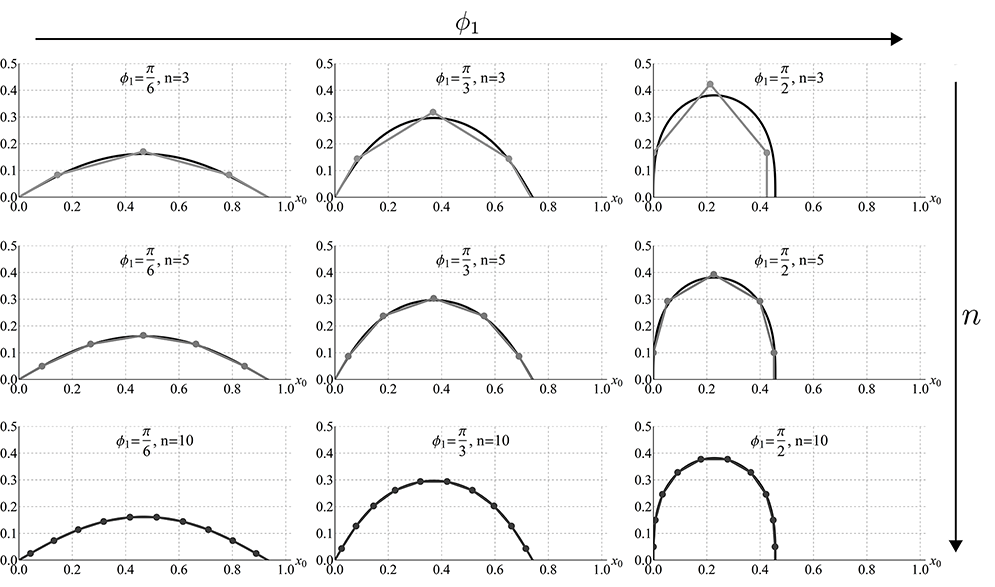}
\caption{Progressively deformed configurations of the discrete structure under compression for increasing values of initial rotation ($\phi_1=\phi(0)=\pi/6,\pi/3,\pi/2$, from left to right), with increasing number of modular elements ($n=3,5,10$, from top to bottom), in comparison with the Elastica continuum limit. Grey dots and segments denote deformed configurations of the discrete systems while black solid lines indicate the corresponding ones of the continuous (Elastica).}
\label{fig.mapping}
\end{figure}

As for the case of tension, the kinematics of the structure subject to compressive axial load can be described in light of the theory of (first order) SDs. Specifically, an approximating sequence properly describing the geometry that characterizes the deformation in this case (Figure \ref{fig.1D_structure}\hyperref[fig.1D_structure]{B}) can be written as follows:  
\begin{equation}
\label{appfunccontraction}
\begin{split}
\textbf{g}_n^h \left( \textbf{x}_0 \right) = H(a_n-1-h) \left[ \left( x_0-c_n \right) \cos\phi_{h+1}+ L_n \left(e_n \cos\phi_1+ \sum_{i=2}^h \cos\phi_i\right)\right] + \\ 
b_n \delta\left(h-a_n\right)\left[ x_0-c_n+L_n \left(e_n \cos\phi_1+ \sum_{i=2}^{a_n} \cos\phi_i\right)\right] +\\
+H(h-a_n-b_n)\left[ \left( x_0-p_n\right) \cos\phi_{n-h+1} +L_n \left( b_n+f_n\cos\phi_1+ 2 \sum_{i=2}^{a_n}\cos \phi_i - \sum_{i=2}^{n-h+1}\cos\phi_i \right)\right]  \hat{\textbf{e}}_1 +\\
+ H(a_n-1-h) \left[ \left( x_0-c_n \right) \sin\phi_{h+1}+ L_n \left(e_n \sin\phi_1+ \sum_{i=2}^h \sin\phi_i\right)\right] + \\ 
+b_n \delta\left(h-a_n\right) \left[ L_n \left[\dfrac{1}{2} \sin\phi_1+ \sum_{i=2}^{a_n} \sin\phi_i\right]\right] +\\
+H(h-a_n-b_n)\left[ \left(p_n -x_0\right) \sin\phi_{n-h+1} +L_n \left(t_n\sin\phi_1+\sum_{i=2}^{n-h+1}\sin\phi_i \right)\right]  \hat{\textbf{e}}_2\,,\\
\forall\,\,x_0 \, \in \, \left] c_n , p_n+2L_n t_n \right[ \,\,,
\end{split} 
\end{equation}
where $H(h-p)$ is the Heaviside function taking value $1$ when $h\geq p$ and $0$ otherwise, and the functions $c_n$, $p_n$, $e_n$, $f_n$ and $t_n$ are introduced, for sake of simplicity, as $c_n=L_n \left(2h-1+\delta(h) \right)/2$, $p_n=L_n \left( 2h-1+\delta(h-n)\right)/2$, $e_n=\left( 1-\delta(h)\right)/2$, $f_n=\left( 1+\delta(h-n)\right)/2$ and $t_n=\left( 1-\delta(h-n)\right)/2$. Consequently, the related gradients are  
\begin{equation}
\label{approxgrad}
\begin{split}
\nabla\textbf{g}_n^h = \left[ H(a_n-1-h) \cos\phi_{h+1}+b_n \delta(h-a_n)+H(h-a_n-b_n)\cos\phi_{n-h+1}\right]  \,\hat{\textbf{e}}_1\otimes\hat{\textbf{e}}_1+\\
+\left[ H(a_n-1-h) \sin\phi_{h+1}-H(h-a_n-b_n)\sin\phi_{n-h+1}\right] \,\hat{\textbf{e}}_2\otimes\hat{\textbf{e}}_1\,,\\
\forall\,\,x_0 \, \in \,\left]  c_n , p_n+2L_n t_n \right[ \,\,.
\end{split}
\end{equation}
By considering that the approximating sequence in \eqref{appfunccontraction} represents the solution of the problem at hand, described by the system in \eqref{discelast}, and by recalling that such a \textit{discrete} problem tends to the differential Euler's Elastica as $n\rightarrow\infty$, it can be deduced that --in the limit $n\rightarrow\infty$-- the approximating sequence above also tends to the solution of the Euler's Elastica, i.e. to the well-known macroscopic deformation describing its deformed shape: 
\begin{equation}
\label{elastmap}
\begin{split}
\textbf{g}(\textbf{x}_0)=\left\lbrace -x_0+\dfrac{2}{\Lambda}\left[ E(am(x_0 \Lambda+K(\sin\frac{\phi_1}{2}),\sin\frac{\phi_1}{2}),\sin\frac{\phi_1}{2})-E(am(K(\sin\frac{\phi_1}{2}),\sin\frac{\phi_1}{2}),\sin\frac{\phi_1}{2})\right] \right\rbrace\, \hat{\textbf{e}}_1 +\\
-\dfrac{2}{\Lambda}\sin\frac{\phi_1}{2}\, cn(x_0 \Lambda+K(\sin\frac{\phi_1}{2}))\,\hat{\textbf{e}}_2\,,
\end{split}
\end{equation}
where $\Lambda=\sqrt{F/B}=2K(\sin(\phi_1/2))/L$, $K(\bullet,\sin(\phi_1/2))$ and $E(\bullet,\sin(\phi_1/2))$ are the elliptic integrals of the first and second kind, respectively, while $am(\bullet,\sin(\phi_1/2))$ and $cn(\bullet,\sin(\phi_1/2))$ are the Jacobi amplitude and the Jacobi cosine amplitude functions. This is graphically illustrated in Figure \ref{fig.mapping}, where the deformed configuration of the structure is plotted for an increasing number of modular elements in comparison with the Elastica (the continuum limit), for different values of the initial rotation.
On the other hand, with regard to the deformation without disarrangements $\textbf{G}$, one obtains from eq. \eqref{approxgrad} that
\begin{equation}
\textbf{G}=\lim_{n \to \infty} \nabla\textbf{g}_n^h=\cos\phi(\textbf{x}_0)\, \hat{\textbf{e}}_1\otimes\hat{\textbf{e}}_1 + \sin\phi(\textbf{x}_0)\, \hat{\textbf{e}}_2\otimes\hat{\textbf{e}}_1\,,
\end{equation}
which exactly coincides with the gradient $\nabla\textbf{g}$ of the macroscopic deformation $\textbf{g}$ in \eqref{elastmap}, thus showing that the considered SD is, in fact, a \textit{classical} deformation, with no sub-macroscopic disarrangements. 
In compliance with the absence of such disarrangements, the tensor $\textbf{M}$ introduced in \eqref{M}
turns out to be equal to $\mathbf{0}$, while the tensor $\textbf{K}$, accounting for the purely sub-macroscopic quote of a first order SD, is the identity tensor and, hence, the virgin configuration coincides with the reference configuration. As a consequence, the stress due to disarrangements $\textbf{S}_d$ vanishes and the component of stress without disarrangements $\textbf{S}_{\backslash}$ is exactly equal to the first Piola-Kirchhoff stress tensor $\textbf{S}=F/A\,\hat{\textbf{e}}_1\otimes\hat{\textbf{e}}_1$.\\
In conclusions, it is important to note that, while disarrangements arising from discontinuities in the sequence $\textbf{g}^h_n$ of deformations are absent in this case, the sequence of the gradients $\nabla \textbf{g}^h_n$ in \eqref{approxgrad} is characterized by discontinuities. This is also shown in Figures (\ref{fig.experiment}) and (\ref{fig.mapping}) where, obviously, slopes of neighboring modules of the compressed discrete structure exhibit different constant slopes as the whole system contracts, this implying that second gradients $\nabla\nabla \textbf{g}^h_n$ are identically zero for the case at hand. These properties of approximating sequences render this geometry a special case of coherent sub-macroscopically affine motions, analyzed in \cite{Owen-2017} for studying elasticity with gradient disarrangements.

\section{Discussion and Conclusions}
In the present work, the nonlinear elastic response of a one-dimensional multi-modular structure under compressive/tensile dead loads has been analyzed, both for a finite number of constituents (discrete structure) and infinite elemental units (continuum structure), in the latter case providing, for the first time, a 1D paradigm of the SDs theory and demonstrating the need to invoke this approach for naturally obtaining augmented hyperelastic models.
\begin{figure}[!h]
\centering
\includegraphics[width=1 \textwidth]{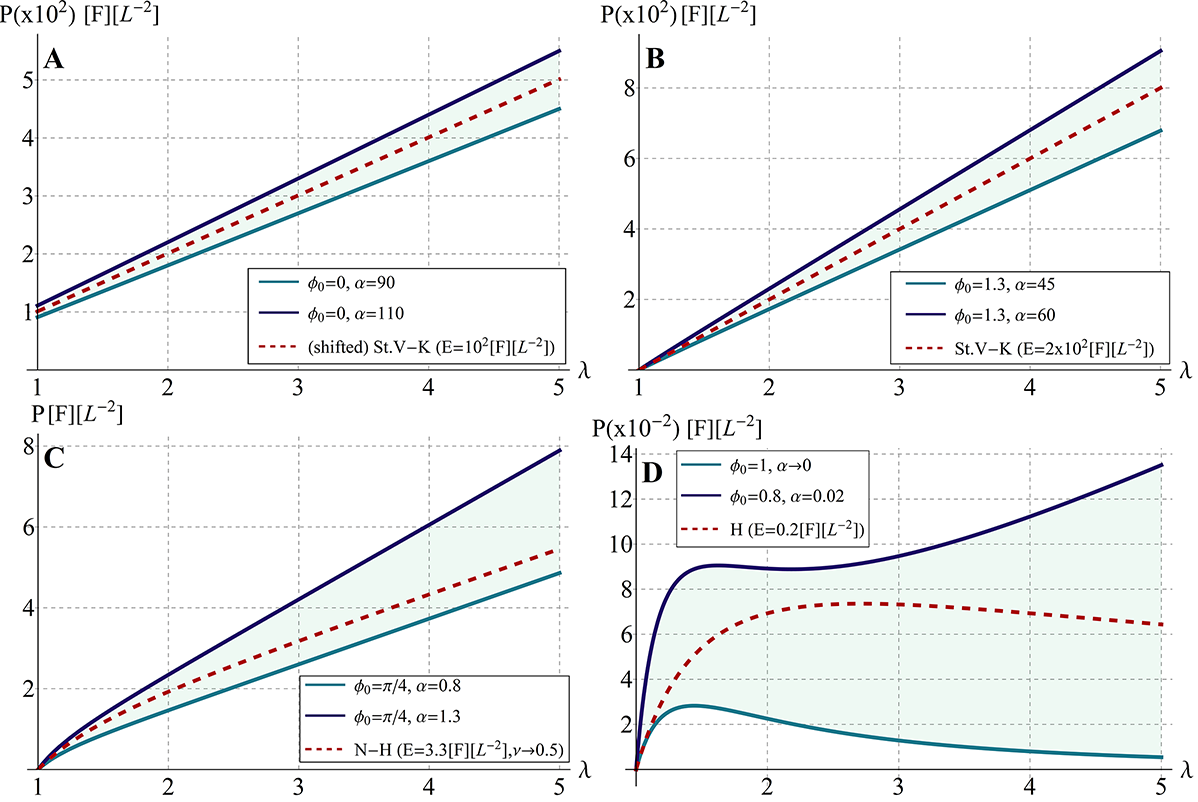} 
\caption{Comparison (in terms of first Piola-Kirchhoff stress $P$ \textit{versus} axial stretch $\lambda$) between standard hyperelastic models (dashed lines) and the macroscopic responses given by the 1D structure studied in the present work --in the \textit{continuum} limit of $n\rightarrow\infty$-- by properly setting the values of its internal parameters $\phi_0$ (initial imperfection) and $\alpha$ (ratio between translational and rotational springs stiffnesses): \textbf{A)} \textit{shifted} Saint Venant-Kirchhoff model (St.V-K), \textbf{B)} Saint Venant-Kirchhoff model, \textbf{C)} neo-Hookean model (N-H) and \textbf{D)} Hencky's model (H), with the upper curve (in blue) a typical balloon-like trend.}
\label{fig.constmodels}
\end{figure}

In particular, with regard to the compressive loading, it has been found that the first bifurcation mode exhibited by the structure replicates, in a discrete form, the solution of the \textit{Euler's Elastica} problem, converging to the classical Elastica in the continuum limit. On the other hand, the behavior of the structure under tensile load --that can be further enriched with the introduction of an initial imperfection-- results to be influenced both by the stiffness of the springs associated with the internal sliders (i.e. by their dependence on the number of elemental moduli) and by the ratio between the values of such stiffness and of the rotational spring at the left end. It has to be highlighted that, in the compressive case, the description in terms of SDs reveals that, as $n\rightarrow \infty$, the relative rotations at the nodes endowed with elastic springs tend to vanish, thus reproducing the classical deformation mode of the continuum without disarrangements. On the contrary, this does not occur under tensile load, where the associated deformation converges to a macroscopic (threshold-activated, for the case of a perfect system) uni-axial elongation, storing sub-macroscopic disarrangements which a standard continuum theory would not have traced.

Additionally, it is interesting to note the plurality of behaviors that the structure provides under tensile load (see Figures \ref{fig.FphiFcr1}-\ref{fig.F2tensimperf}), by tuning internal properties such as the ratio between the springs stiffnesses and/or the initial imperfection. In fact, by modulating these ratios, various classical nonlinear constitutive models can be derived as the result of a consistent bottom-up procedure incorporating SDs. To show this in detail, we focus on the macroscopic behavior of the 1D continuum structure ($n\rightarrow\infty$) and set the stiffness of the sliders' springs proportional to the number of units $n$. For this case, Figure \ref{fig.constmodels} illustrates that one can reproduce at least four commonly adopted hyperelastic constitutive models, therein plotted in terms of first Piola-Kirchhoff stress versus axial (macroscopic) stretch. In particular, if the system is \textit{perfect} ($\phi_0=0$), one finds a \textit{threshold-activated} Saint Venant-Kirchhoff model \cite{Bigoni2012} (written with reference to the Cauchy strain measure derived from the Seth-Hill formula \cite{Bigoni2012,Hill_1968}, for small strain giving the classic linear Hooke's law \cite{timoshenko1967}). On the other hand, in the case of \textit{imperfect} system, by properly playing on the microstructural parameters $\phi_0$ and $\alpha$, one can obtain a wider class of hyperelastic curves including the neo-Hookean law \cite{holzapfel2000} and responses providing softening such as the Hencky's model \cite{anand1979,xiao2002} and the non-monotonic behavior recognized in inflated balloons, in terms of pressure-radius relation \cite{Muller1996}.
\begin{figure}[!h]
\centering
\includegraphics[width=.9 \textwidth]{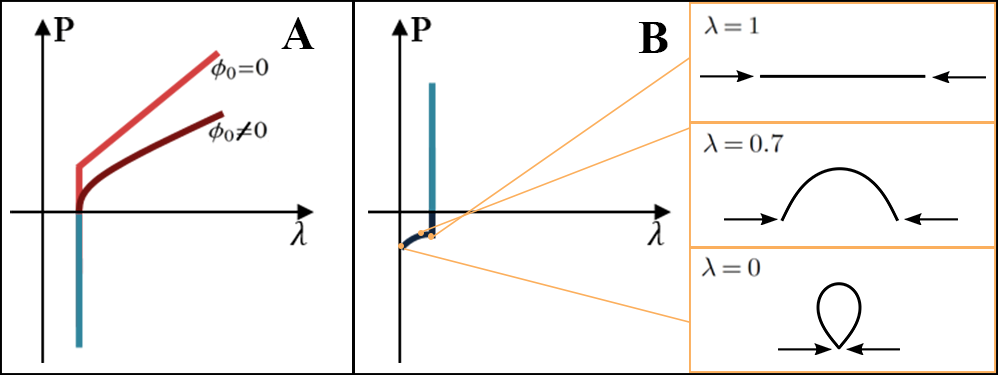} 
\caption{Whole (i.e. both under compressive and tensile regime) \textit{macroscopic} response of the structure in terms of first Piola-Kirchhoff stress \textit{versus} axial stretch, for the two complementary constitutive assumptions made in the sections above: \textbf{A)} structure deforming only under tensile load --with or without a threshold behavior according to the absence or presence of sub-macroscopic imperfections, respectively-- offering infinite stiffness to compression ($k\rightarrow\infty$); \textbf{B)} structure deforming only under compressive load, offering infinite stiffness to tension ($K_{\Delta}\rightarrow\infty$). In the latter case, the apparently inconsistent finite value of the stress as the macro-stretch $\lambda$ goes to zero –-a condition unseen in standard continua-– is here in principle admissible as a consequence of the fact that such stretch measures the actual distance between the 1D element endpoints, that in case of compression could coincide –-or even invert their position so producing negative stretches-- for extremely large deformation of the elastica. In this regard, the inset on the right side reports illustrative sketches of the deformed macroscopic system for decreasing value of $\lambda$.}
\label{fig.3sketch}
\end{figure}
Figure \ref{fig.3sketch} then combines the responses of the multi-modular structure under the examined loading condition types, that is for tensile and compressive external applied forces, accounting for the complementary constitutive assumptions respectively made in the two cases in the previous sections. In particular, Figure \ref{fig.3sketch}\hyperref[fig.3sketch]{A} shows the response of a structure which can deform only under tensile load --with or without a threshold behavior, according to the absence or to the presence of sub-macroscopic imperfections, respectively-- offering instead infinite stiffness to compression ($k\rightarrow\infty$). On the other hand, Figure \ref{fig.3sketch}\hyperref[fig.3sketch]{B} illustrates the complementary case of a system which undergoes deformation only under compressive load, being infinitely stiff to tension ($K_{\Delta}\rightarrow\infty$).\\
In conclusions, it is foreseen that the presented SDs-based strategy could be in future generalized to two- and three-dimensional structures, in this way widening the range of applicable loading conditions and the richness of the potential resulting microstructural and global mechanical behaviors. Moreover, it is envisaged that dissipation --irreversible and time-dependent-- phenomena could be also involved, for example through the integration of dashpot elements, thus possibly including the chance of reproducing classical models of linear viscoelasticity (e.g. Maxwell as well as Kelvin-Voigt models). In this way, valuable homogenized models could be then obtained, which would allow to both interpret complex macroscopic behaviors of hierarchically organized materials and biological tissues and trace back some relevant physical phenomena by means of \textit{localization} procedures.

\paragraph{Ethics.}The authors declare that all the ethical issues were respected.

\paragraph{Data Accessibility.} This article has no additional data. However, further data can be attained by replacing other parameters in the proposed model.

\paragraph{Authors’ Contributions.}All the Authors equally contributed to the paper. In particular: S.P. developed the model, performed analysis and wrote the paper; M.F. conceived the idea, advised the model development, wrote and reviewed the paper, with L.D. and S.P. translating the results in terms of Structured Deformations; L.D. and D.O. provided the theory of Structured Deformations, reviewed and edited the final draft.

\paragraph{Competing Interests.}We have no competing interests.

\paragraph{Funding.}The work was supported by the grant from Italian Ministry of Education, Universities and Research (MIUR) ARS01-01384-PROSCAN, by the grant from University of Napoli "Federico II" E62F17000200001-NAPARIS and partially supported by the grant ERC-2013-ADG-340561-INSTABILITIES.

\paragraph{Acknowledgements.}MF gratefully acknowledges the support of the grants by MIUR (PROSCAN) and by the University of Napoli "Federico II" (NAPARIS). LD gratefully acknowledges the partial support of the grant ERC-2013-ADG-340561-INSTABILITIES.

\newpage
\bibliographystyle{unsrt}  

\end{document}